\documentclass[A4paper,11pt]{article}
\usepackage{microtype}%if unwanted, comment out or use option "draft"
\usepackage{lineno}
\usepackage{epsfig}
\usepackage{microtype}
\usepackage{multirow}
\usepackage{graphicx}
\usepackage{enumitem}
\usepackage[linesnumbered,ruled, lined]{algorithm2e}
\usepackage{algpseudocode}
\usepackage{epsfig,enumerate,amsmath,amsfonts,amssymb,amsthm,mathrsfs,ifpdf}
\usepackage{indentfirst,relsize}
\usepackage[numbers]{natbib}
\usepackage{setspace}
\usepackage{enumerate}
\usepackage{latexsym}
\usepackage{stackrel}
\usepackage[all]{xy}
\usepackage[bottom=2cm,top=2cm,left=2cm,right=2cm]{geometry}
\usepackage[usenames,dvipsnames]{pstricks}
\usepackage{pst-grad} 
\usepackage{pst-plot} 
\usepackage{xspace}
\usepackage{hyperref}

\theoremstyle{plain}
\newtheorem{theo}{Theorem}[section]
\newtheorem{lem}[theo]{Lemma}

\newtheorem{coro}[theo]{Corollary}

\newtheorem{cl}[theo]{Claim}

\newtheorem{pre}[theo]{Proposition}
\theoremstyle{definition}
\newtheorem{defi}[theo]{Definition}
\newtheorem{rem}{Remark}
\newtheorem{obs}[theo]{Observation}

\newtheorem*{obs*}{Observation}

\newcommand{\iporacle}{$\mbox{{\sc IP}}$\xspace}
\newcommand{\iporacledec}{$\mbox{{\sc Dec-IP}}$\xspace}
\newcommand{\element}{$\mbox{{\sc ME}}$\xspace}

\newcommand{\vmv}{$\mbox{{\sc BF}}$\xspace}
\newcommand{\vmvdec}{$\mbox{{\sc Dec-BF}}$\xspace}

\newcommand{\dpd}[2]{\langle {#1}, %\cdot 
#2\rangle}
  
\newcommand{\distsymmmat}{\mbox{{\sc Dist-Symm-Matrix}}\xspace}
\newcommand{\distsymmmatguess}{\mbox{{\sc Dist-Symm-Matrix-Guess}}\xspace}
\newcommand{\distrows}{\mbox{{\sc Dist-Bet-Rows}}\xspace}
\newcommand{\resdistrows}{\mbox{{\sc Restrict-Dist-Bet-Rows}}\xspace}
\newcommand{\randsamp}{\mbox{{\sc Approx-Sample}}\xspace}
\newcommand{\dist}{{{\bf D}}_{\bf M} \xspace}

\newcommand{\ham}{\mbox{{\bf d}}_{{{\bf H}}}}
\newcommand{\ident}{\mbox{{\sc Identity}}}

\newcommand{\eps}{\vareps}
\newcommand{\uta}{{{\bf \Delta^A}}}
\newcommand{\lta}{{{\bf \Delta_A}}}
\newcommand{\utb}{{{\bf \Delta^B}}}
\newcommand{\ltb}{{{\bf \Delta_B}}}
\newcommand{\bx}{{\bf x}\xspace}
\newcommand{\by}{{\bf y}\xspace}

\newcommand{\disj}{\mbox{{\sc Disjointness}}}
\newcommand{\high}{1-\frac{1}{\poly{n}}}
\newcommand{\mtx}[1]{{\bf #1}}
\newcommand{\poly}[1]{\textit{poly}\left( #1 \right)}

\newcommand{\tTh}{\widetilde{\Theta} \xspace}

\renewcommand{\hat}{\widehat}
\renewcommand{\O}{\mathbb{O}}

\newcommand{\size}[1]{\left| #1 \right|}

 \newcommand{\nbuck}{ \log _{\eps/50} n }

\newcommand{\E}{\mathbb{E}}
\newcommand{\remove}[1]{}

\newcommand{\R}{\mathbb{R}}
\newcommand{\N}{\mathbb{N}}

\newcommand{\cA}{\mathcal{A}}

\newcommand{\cE}{\mathcal{E}}

\newcommand{\cP}{\mathcal{P}}
\newcommand{\cV}{\mathcal{V}}

\newcommand{\Oh}{\mathcal{O}}

\newcommand{\tOh}{\widetilde{{\mathcal O}}}

\newcommand{\cH}{\mathcal{H}}

\newcommand{\Z}{\mathbb{Z}}
\newcommand{\mat}{\mtx}

\newcommand{\vareps}{\epsilon}
\newcommand{\pr}{\mathbb{P}}

\newcommand{\complain}[1]{\textcolor{red}{#1}}
\newcommand{\comments}[1]{\textcolor{blue}{\bf{#1}}}

\begin{document}
\title{Distance Estimation Between Unknown Matrices Using Sublinear Projections on Hamming Cube}
\remove{\title{
Sublinear time algorithm for Hamming Distance between Matrices using Projections \\
{Sub-linear Random Projections suffice to Estimate Distance between
  Binary Images}\\
{Estimating the Hamming Distance between Unknown Matrices using Projections on Hamming Vectors}\\
{Sublinear algorithm for estimating Hamming distance between  matrices using aggregate query}
}}

\date{}
%\author{}

\author{
Arijit Bishnu\footnote{Indian Statistical Institute, Kolkata, India}
\and 
Arijit Ghosh\footnotemark[1]
\and 
Gopinath Mishra\footnotemark[1]
}

\maketitle
%\vspace{-1.5cm}
\begin{abstract}
\noindent
Using geometric techniques like projection and dimensionality reduction, we show that there exists a randomized sub-linear time algorithm that can estimate the Hamming distance between two matrices. Consider two matrices ${\bf A}$ and ${\bf B}$ of size $n \times n$ whose dimensions are known to the algorithm but the entries are not. The entries of the matrix are real numbers.
The access to any matrix is through an oracle that computes the projection of a row (or a column) of the matrix on a vector in $\{0,1\}^n$. 
We call this query oracle to be an {\sc Inner Product} oracle (shortened as {\sc IP}).
We show that our algorithm returns a $(1\pm \epsilon)$ approximation to ${{\bf D}}_{\bf M} ({\bf A},{\bf B})$ with high probability by making ${\cal O}\left(\frac{n}{\sqrt{{{\bf D}}_{\bf M} ({\bf A},{\bf B})}}\mbox{poly}\left(\log n, \frac{1}{\epsilon}\right)\right)$ oracle queries, where ${{\bf D}}_{\bf M} ({\bf A},{\bf B})$ denotes the Hamming distance (the number of corresponding entries in which ${\bf A}$ and ${\bf B}$ differ) between two matrices ${\bf A}$ and ${\bf B}$ of size $n \times n$.  We also show a matching lower bound on the number of such {\sc IP} queries needed. Though our main result is on estimating ${{\bf D}}_{\bf M} ({\bf A},{\bf B})$ using {\sc IP}, we also compare our results with other query models.
\end{abstract}

\section{Introduction}\label{sec:intro}
\noindent
Measuring similarity between entities using a distance function has been a major area of focus in computer science in general and computational geometry in particular~\cite{10.1145/1824777.1824791,10.1145/3185466,DBLP:journals/dcg/DriemelHW12,DBLP:journals/siamcomp/DriemelH13,DBLP:journals/siamcomp/AgarwalAKS14}. Distance computations require access to the entire data and thus can not escape computations that are linear in time complexity. In this era of big data, seeing the entire data may be too much of an ask and trading precision for a time efficient algorithm is a vibrant area of study in property testing~\cite{G2017}. Testing properties of binary images with sub-linear time algorithms has been a focus of property testing algorithms~\cite{talg/RonT14,random/Raskhodnikova03,icalp/BermanMR16,pami/KleinerKNB11,compgeom/BermanMR16}. Matrices are ubiquitous in the sense that they represent or abstract a whole gamut of structures like adjacency matrices of geometric graphs and visibility graphs, images, experimental data involving 0-1 outcomes, etc. Pairwise distance computations between such matrices is a much needed programming primitive in image processing and computer vision applications so much so that the widely used commercial toolbox MATLAB of MathWorks\textregistered ~\cite{matlab} has an inbuilt function call named \texttt{pdist2($\cdot,\cdot,\cdot$)}~\cite{matlab-pdist} for it. Other open source based software packages also have similar primitives~\cite{scilab-pdist}. For all these primitives, the matrices need to be known. But in all situations where access to the matrices are restricted (say, because of security, privacy or communication issues) except for an oracle access, we want to know how much the two matrices differ in their entries. Keeping in line with the above, we focus on distance estimation problem between two matrices whose dimensions are known to the algorithm but the entries are unknown; the access to the matrices will be through an oracle.  This oracle, though linear algebraic in flavor, has a geometric connotation to it. We hold back the discussion on the motivation of the oracle till Section~\ref{ssec:oracle-def}.

\paragraph*{Notations} 
\noindent 
In this paper, we denote the set $\{1,\ldots,t\}$ by $[t]$ and $\{0,\ldots,t\}$ by $[[t]]$. 
 \remove{For a non-empty set $X$ and a given parameter $\eps \in (0,1)$, an \emph{almost uniform} sample of $X$ means each element of $X$ is sampled with probability values that lie in the interval $\left[(1-\eps)\frac{1}{\size{X}},(1+\eps)\frac{1}{\size{X}} \right]$.} 
 For a matrix $\mtx{A}$, $\mtx{A}(i,j)$ denotes the element in the $i$-th row and $j$-th column of $\mtx{A}$. Unless stated otherwise, $\mtx{A}$ will be a  matrix with real entries. $\mtx{A}(i,*)$ and $\mtx{A}(*,j)$ denote the $i$-th row vector and $j$-th column vector of the matrix $\mtx{A}$, respectively. \remove{$A \in 
[[\rho]]^{n \times n}$ means $A_{ij} \in [[\rho]]$ for each $i,j \in [n]$, {$\rho \in \N$}.} Throughout this paper, the number of rows or columns of a square matrix $\mtx{A}$ will be $n$.
Vectors are matrices of order $n \times 1$ and will be represented using bold face letters. Without loss of generality, we consider $n$ to be a power of $2$. The $i$-th coordinate of a vector ${\bf x}$ will be denoted by 
$x _{i}$. We will denote by $\bf{1}$ the vector with all coordinates $1$. Let $\{0,1\}^n$ denote the set of $n$-dimensional vectors with entries either $0$ or $1$. \remove{For $\bx \in \R^{n}$, ${\bf 1}_{\bx}$ is a vector in $\{0,1\}^{n}$ whose $i$-th coordinate is $1$ if $x_{i} \neq 0$ and $0$ otherwise; $\mathrm{nnz}(\bx) = \size{ i \in [n]: \; x_{i} \neq 0 }$ denotes the number of non-zero components of the vector.} 
By $\dpd{\bx}{\by}$, we denote the standard 
inner product of $\bx$ and $\by$, that is, $\dpd{\bx}{\by}=\sum_{i=1}^n x_iy_i$. 
$P$ is a $(1 \pm \eps$)-approximation to $Q$ means $\size{P-Q} \leq \eps \cdot Q$. \remove{
$\tOh(f)-$ $\Oh(f)$ hiding a term polynomial in $\log n$ and $\frac{1}{\eps}$.} The statement \emph{with high probability} means that the probability of success is at least $1-\frac{1}{n^c}$, where $c$ is a positive constant. $\widetilde{\Theta}(\cdot)$ and $\widetilde{\Oh}(\cdot)$ hides a $\mbox{poly}\left(\log n, \frac{1}{\eps}\right)$ term in the upper bound. {$|| \cdot ||_p$ denotes the usual $\ell_p$ distance.}

\subsection{Query oracle definition and motivation, problem statements and our results}
\label{ssec:oracle-def}
\begin{defi}[{Matrix distance}]
\remove{Let $\mtx{A}$ and $\mtx{B}$ be two $n\times n$ matrices.} The \emph{matrix-distance} between two matrices $\bf{A}$ and $\bf{B}$ of size $n \times n$ is the number of pairwise mismatches and is denoted and defined as
$$\dist(\mtx{A},\mtx{B})=\size{\{(i,j):i,j\in [n], \mtx{A}(i,j)\neq \mtx{B}(i,j)\}}.$$ 
\end{defi}
As alluded to earlier, the matrices can not be accessed directly, the sizes of the matrices are known but the entries are unknown. We will refer to the problem as the \emph{matrix distance} problem. We consider the following query models to solve the matrix distance problem in this paper. 
\paragraph*{Query oracles for unknown matrix $\mat{A} \in \R^{n \times n}$:} The main query oracle access used in this work is based on the inner product of two vectors and is defined as follows:
\begin{description}
 \item[\bf {\sc Inner Product} (\iporacle):] Given a row index $i \in [n]$ (or, a column index $j \in [n]$) and a vector ${\bf{v}} \in \{0,1\}^{n}$, the \iporacle query to $\mtx{A}$ reports the value of $\dpd{\mat{A}(i,*)}{\bf{v}}$ ($\dpd{\mat{A}(*,j)}{\bf{v}}$). If the input index is for row (column), we refer the corresponding query as row (column) \iporacle query.
\end{description}

This linear algebraic oracle access has a geometric connotation to it in terms of \emph{projection} onto Hamming vectors -- we exploit this understanding in our work. This oracle access is also motivated from a practical angle. A dot product operation is a fit case for parallelization using a {\sc Single Instruction Multiple Data (SIMD)} architecture~\cite{compu-arch-book}. Modern day {\sc GPU} processors provide instruction level parallelism. In effect, {\sc NVIDIA} GPUs that are built on {\sc CUDA} architecture, provide dot product between two vectors as a single API call~\cite{sanders10,cuda-dot-nvidia}. Thus, there exists practical implementation of the query oracle access that we use. There are also examples of programming languages supporting SIMD intrinsics that can compute dot product~\cite{simd-intrinsics}. There is a caveat though -- in terms of resource, more processors are used. For us, in this work, the query complexity is the number of calls to the \iporacle. As mentioned, modern day architectures allow us to convert each \iporacle query to a one cycle computation with more processors used in parallel.

In the power hierarachy of matrix based query oracles, \iporacle surely wields some power vis-a-vis solving certain problems~\cite{BishnuGMP2019}~\footnote{{But the \iporacle{} defined in this paper is weaker than that defined in~\cite{BishnuGMP2019} -- in their case, one is allowed to query for inner product of rows/columns of matrices with vectors in $\R^n$.}}. An obvious question that confronts an algorithm designer is whether a weaker oracle can do the same job at hand (here, computing matrix distance). With that in mind, we define the following two oracles and show that their query complexity lower bounds on the matrix distance problem match the trivial upper bounds. That shows the justification for use of \iporacle.
\begin{description}
\item[\bf {\sc Matrix Element} (\element):] Given two indices $i,j \in [n]$, the \element query to $\mat{A}$ returns the value of $\mat{A}(i,j)$.
\item[\bf {\sc Decision Inner Product} (\iporacledec):] Given a row index $i \in [n]$ (or, a column index $j \in [n]$) and a vector ${\bf{v}} \in \{0,1\}^{n}$, the \iporacledec query to $\mtx{A}$ reports whether $\dpd{\mat{A}(i,*)}{\bf{v}}$ ($\dpd{\mat{A}(*,j)}{\bf{v}}$) $=0$. If the input index is for row (column), we refer the corresponding query as row (column) \iporacledec query.
\remove{
\item[\bf {\sc Inner Product} (\iporacle):] Given a row index $i \in [n]$ (or, a column index $j \in [n]$) and a vector ${\bf{v}} \in \{0,1\}^{n}$, the \iporacle query to $\mtx{A}$ reports the value of $\dpd{\mat{A}(i,*)}{\bf{v}}$ ($\dpd{\mat{A}(*,j)}{\bf{v}}$). If the input index is for row (column), we refer the corresponding query as row (column) \iporacle query.
}
\remove{\item[\bf {\sc Decision Bilinear Form} (\vmvdec):] Given two vectors ${\bf x}, {\bf y} \in \R ^n$, the \vmvdec query to $\mtx{A}$ reports whether ${\bf x}^T \mat{A}{\bf y}=0$. 
\item[\bf {\sc Bilinear Form} (\vmv):] Given two vectors ${\bf x}, {\bf y} \in \R ^n$, the \vmv query to $\mtx{A}$ reports the value of ${\bf x}^T \mat{A}{\bf y}$. }
\end{description}
The following remark highlights the relative power of the query oracles.
\begin{rem}\label{rem:power}
Each \element query can be simulated by using one \iporacledec{} oracle, and each \iporacledec{} oracle can be simulated by using one \iporacle{} query.  
\end{rem}

\paragraph*{Our results} Our main result is an algorithm for estimating the distances between two unknown matrices using \iporacle{}, and the result is formally stated as follows. Unless otherwise mentioned, all our algorithms are randomized.

\begin{theo}[{\bf Main result: Estimating the distance between two arbitrary matrices}]\label{theo:matrix-dist}
There exists an algorithm that has \iporacle query oracle access to unknown matrices $\mtx{A}$ and $\mtx{B}$, takes an $\eps\in (0,1)$ as an input, and returns a $(1\pm \eps)$ approximation to $\dist (\mtx{A},\mtx{B})$ with high probability, and makes 
$\Oh\left( \left({n}/\sqrt{\dist (\mtx{A},\mtx{B})}\right) \mbox{poly}\left(\log n, \frac{1}{\eps}\right)\right)$ \iporacle queries. 
\end{theo}
We also show that our algorithm (corresponding to the above theorem) is optimal, if we ignore the $\mbox{poly}\left(\log n, \frac{1}{\eps}\right)$ term, by showing (in Theorem~\ref{theo:lb-main}) that any algorithm that estimates $\dist(\mtx{A},\mtx{B})$ requires $\Omega\left({n}/{\sqrt{\dist(\mtx{A},\mtx{B})}}\right)$ \iporacle queries. For the sake of completeness in understanding the power of \iporacle, we study the matrix distance problem also using two weaker oracle access -- \element and \iporacledec. Our results are summarized in Table~\ref{table:results-conflict-and-est-graph} and they involve both upper and almost matching lower bounds in terms of the number of queries needed. Note that all of our lower bounds hold even if one matrix (say $\mat{A}$) is known and both matrices ($\mat{A}$ and $\mat{B}$) are symmetric binary matrices.

\begin{table}[thb]
\centering
\begin{tabular}{|c || c | c | c |} 

 %\hline
 \hline \hline
  Query Oracle & \element & \iporacledec & \iporacle \\
 % \multirow{2}{*}{power, if any} & \multicolumn{2}{c|}{} & \multicolumn{2}{c}{} \\
 \cline{1-4}
  %& \conflictest & \colorverify & \conflictest & \colorverify \\
       \hline \hline
  \multirow{2}{*} {Upper Bound} & $\tOh \left(\frac{n^2}{D}\right)$   & $\tOh \left(\frac{n^2}{D}\right)$ & $\tOh \left(\frac{n}{\sqrt{D}}\right)$\\
  & (Trivial) & (Trivial) & (Theorem~\ref{theo:matrix-dist}) \\
  \cline{1-4}
  \multirow{2}{*} {Lower Bound}  & $\Omega \left(\frac{n^2}{D}\right)$ &  $\Omega \left(\frac{n^2}{D}\right)$ & $\Omega \left(\frac{n}{\sqrt{D}}\right)$\\
  & (Corollary~\ref{coro:lb-me} )  & (Theorem~\ref{theo:lb-main} ) & (Theorem~\ref{theo:lb-decip})  \\
   \hline \hline
  
\end{tabular}
\caption{Our results. In this table, $D=\dist (\mtx{A},\mtx{B})$.}
 
\label{table:results-conflict-and-est-graph}
\end{table}

\remove{
To justify, the consideration for \iporacle, we also prove (in Theorem~\ref{theo:lb-decip}) a lower bound of  $\Omega\left(\frac{n^2}{{\dist(\mtx{A},\mtx{B})}}\right)$ results for estimating $\dist(\mtx{A},\mtx{B})$ distance between two matrices when we have \iporacledec{} query access to the matrices. Note that, by Remark~\ref{rem:power}, the same lower bound holds if we have \element query access to the matrices (as each \element{} query can be simulated by using a \iporacledec query oracle). The lower bound of $\Omega\left(\frac{n}{\sqrt{\dist(\mtx{A},\mtx{B})}}\right)$ is essentially tight when we have \element (\iporacledec) query oracle access to the matrices.} 
%\complain{[Need to check the theorem statement later.]}

\begin{rem}\label{rem:plus-minus}
 Note that an \iporacle{} query to a matrix $\mat{A}$ answers inner product of a specified 
row (column) with a given binary vector. However, we will describe subroutines (of the algorithm for estimating the distance between two matrices) that ask for inner product of a specified 
row (column) with a given vector ${\bf r} \in \{-1,1\}^n$. This is not a problem as $
\dpd{\mat{A}(i,*)}{{\bf r}}$ ($\dpd{\mat{A}(*,j)}{{\bf r}}$) can be computed by 
using two \iporacle{} queries (with binary vectors)~\footnote{For ${\bf 
r} \in \{-1,1\}^n$, consider ${\bf v_1}, {\bf v_{-1}} \in \{0,1\}^n$ indicator 
vectors for $+1$ and $-1$ coordinates in ${\bf r} \in \{-1,1\}^n$, respectively. 
Then $\dpd{\mat{A}(i,*)}{{\bf r}}=\dpd{\mat{A}(i,*)}{{\bf v_1}}-\dpd{\mat{A}(i,*)}
{{\bf v_{-1}}}$. So, $\dpd{\mat{A(i,*)}}{{\bf r}}$ can be computed with two 
\iporacle{} queries $\dpd{\mat{A}(i,*)}{{\bf v_1}}$ and $\dpd{\mat{A}(i,*)}{{\bf 
v_{-1}}}$. Similar argument also holds for $\dpd{\mat{A}(*,j)}{{\bf r}}$.}. For simplicity, we refer $
\dpd{\mat{A}(i,*)}{{\bf r}}$ ($\dpd{\mat{A}(*,j)}{{\bf r}}$) also as \iporacle{} query in our algorithm.
\end{rem}

%\complain{Gopi to All: Shall we write here something about inner product with $\R^n$? Need a discussion.}

 \subsection{Related work}
 \noindent
 There are works in property testing and sub-linear geometric algorithms~\cite{ChazelleLM05, CzumajSZ00, CzumajS01, CzumajEFMNRS03}. In the specific problem that we deal with in this paper, to the best of our knowledge, Raskhodnikova~\cite{random/Raskhodnikova03} started the study of property testing of binary images in the \emph{dense image model}, where the number of 1-pixels is $\Omega(n^2)$. The notion of distance between matrices of the same size is defined as the number of pixels (matrix entries) on which they differ. The relative distance is the ratio of the distance and the number of pixels in the image. In this model, Raskhodnikova studies three properties of binary images -- connectivity, convexity, and being a half-plane -- in the property testing framework. Ron and Tsur~\cite{talg/RonT14} studied property testing algorithms in the sparse binary image model (the number of 1-pixels is $O(n)$) for connectivity, convexity, monotonicity, and being a line. The distance measure in this model is defined by the fraction of differing entries taken with respect to the actual number of 1’s in the matrix. As opposed to treating binary images as discrete images represented using pixels as in~\cite{talg/RonT14,random/Raskhodnikova03}, Berman et al.~in \cite{compgeom/BermanMR16} and \cite{rsa/BermanMR19} treated them as continuous images and studied the problem of property testing for convexity of 2-dimensional figures with only uniform and independent samples from the input. To the best of our knowledge, computing distances between binary images has not been dealt with in the sub-linear time framework.  

\paragraph*{Organization of the paper} We prove Theorem~\ref{theo:matrix-dist} (in Section~\ref{sec:arbit}) through a sequence of results. To prove  Theorem~\ref{theo:matrix-dist} that estimates the distance between two arbitrary matrices, we need a result that estimates the distance between two symmetric matrices (Lemma~\ref{lem:matrix-dist-symm} in Section~\ref{sec:dist-symm}) that in turn needs a result on the estimation of the distance between two symmetric matrices with respect to a parameter $T$ (Lemma~\ref{lem:matrix-dist-symm-T} in Section~\ref{sec:dist-symm}). Lemma~\ref{lem:matrix-dist-symm-T} is the main technical lemma that uses dimensionality reduction via Johnson Lindenstrauss lemma crucially. The technical overview including the proof idea of Lemma~\ref{lem:matrix-dist-symm-T} is in Section~\ref{sec:prelim-tech}. The detailed proof idea is in Section~\ref{sec:pf-symm-T}. Using communication complexity, we prove our lower bound results in Section~\ref{sec:lower}.

\remove{\begin{table}[thb]
\centering
\begin{tabular}{||c | c |c|| } 

 \hline
 \hline
 
  Query oracle  & Query complexity & Comments\\
  \hline \hline
  \element & $\tTh \left(\frac{n^2}{D}\right)$ & \\

  \hline
  % \local & $\tTh \left(\frac{n^2}{D}\right)$ \\
  %\hline 
  \iporacledec & $\tTh \left(\frac{n^2}{D}\right)$ & Theorem~\ref{} \\
  \hline
  \iporacle & $\tTh \left(\frac{n}{\sqrt{D}}\right)$ & Theorem~\ref{theo:matrix-dist}\\
  %\hline
  % \vmvdec & $\tTh \left(\frac{n^2}{D}\right)$ \\
  %\hline
 % \vmv & $\tTh(1)$\\
  
  \hline \hline
  
\end{tabular}
\caption{Our results. In this table, $D=\dist (\mtx{A},\mtx{B})$.}
\label{table:results}
\end{table} }

\section{Matrix-Distance between two symmetric matrices}
\label{sec:dist-symm}
\noindent
This Section builds up towards a proof of Theorem~\ref{theo:matrix-dist} by first giving an algorithm that estimates the matrix-distance between two unknown symmetric matrices (instead of arbitrary matrices as in Theorem~\ref{theo:matrix-dist}) with high probability. The result is formally stated in Lemma~\ref{lem:matrix-dist-symm}. In Section~\ref{sec:arbit}, we will discuss how this result (stated in Lemma~\ref{lem:matrix-dist-symm}) can be used to prove Theorem~\ref{theo:matrix-dist}. 
\begin{lem}[{\bf Estimating the distance between two symmetric matrices}]\label{lem:matrix-dist-symm}
There exists an algorithm $\distsymmmat(\mat{A},\mat{B},\vareps)$, described in Algorithm~\ref{algo:dist-wlow}, that has \iporacle query access to unknown {\em symmetric} matrices $\mtx{A}$ and $\mtx{B}$, takes an $\eps \in \left(0,\frac{1}{2}\right)$ as an input, and returns a $(1 \pm \eps)$-approximation to $\dist (\mtx{A},\mtx{B})$ with high probability, making $\tOh\left({n}/{\sqrt{\dist (\mtx{A},\mtx{B})}}\right)$ \iporacle queries.% Note that $\mat{A}=\uta+\lta$ and $\mat{B}=\utb+\ltb$.
\end{lem}

First, we prove a parameterized version of the above lemma in  Lemma~\ref{lem:matrix-dist-symm-T} where we are given a parameter $T$ along with an $\eps \in \left(0,\frac{1}{2}\right)$ and we can obtain an approximation guarantee on $\dist \left(\mat{A},\mat{B}\right)$  as a function of both $T$ and $\eps$. One can think of $T$ as a guess for $\dist(\mtx{A},\mtx{B})$.
\begin{lem}[{\bf Estimating the distance between two symmetric matrices w.r.t.\ a parameter $T$}]\label{lem:matrix-dist-symm-T}
There exists an algorithm $\distsymmmatguess(\mtx{A}, \mtx{B},\vareps,T)$, described in Algorithm~\ref{algo:random-order}, that has \iporacle query access to unknown {\em symmetric} matrices $\mtx{A}$ and $\mtx{B}$, takes parameters $T$ and $\eps \in \left(0,\frac{1}{2}\right)$ as inputs, and returns $\widehat{d}$ satisfying $\left(1-\frac{\eps}{10}\right)\dist \left(\mat{A},\mat{B}\right)-\frac{\eps}{1600}T \leq \hat{d}\leq \left(1+\frac{\eps}{10}\right)\dist \left(\mat{A},\mat{B}\right)$ with high probability, and makes $\tOh\left({n}/{\sqrt{T}}\right)$ queries. Note that here $T$ is at least a suitable polynomial in $\log n$ and ${1}/{\eps}$. 
\end{lem}

 In Section~\ref{sec:prelim-tech}, we discuss some preliminary results to prove Lemma~\ref{lem:matrix-dist-symm-T}. The proof of Lemma~\ref{lem:matrix-dist-symm-T} is given in Section~\ref{sec:pf-symm-T}. If the guess $T \leq \dist(\mtx{A},\mtx{B})$, $\distsymmmatguess(\mat{A},\mat{B},\vareps,T)$ (as stated in Lemma~\ref{lem:matrix-dist-symm-T}) returns 
a $(1\pm \eps)$-approximation to $\dist(\mtx{A},\mtx{B})$ with high probability. However, this dependence on $T$ can be overcome to prove Lemma~\ref{lem:matrix-dist-symm}. We will discuss about the proof of Lemma~\ref{lem:matrix-dist-symm} in Appendix~\ref{sec:pf-symm}.

\subsection{Technical preliminaries to prove Lemma~\ref{lem:matrix-dist-symm-T}}
\label{sec:prelim-tech}
\noindent
%\comments{Now, Observation 2.3 in the previous draft is not needed as it is captured in Remark 2.}
The matrix distance $\dist \left(\mat{A},\mat{B}\right)$ can be expressed in terms of the notion of a distance between a row (column) of matrix $\mat{A}$ and a row (column) of matrix $\mat{B}$ as follows: 
\remove{
Definition~\ref{defi:dist-row} defines the notion of distance between a row (column) of a matrix $\mat{A}$ and a row (column) of a matrix $\mat{B}$. Then Observation~\ref{obs:dist} expresses $\dist \left(\mat{A},\mat{B}\right)$ in terms of the sum of the distances between rows (columns) of $\mat{A}$ and corresponding rows (columns) of $\mat{B}$.
}
\begin{defi}[{\bf Distance between two rows (columns)}]\label{defi:dist-row}
Let $\mtx{A}$ and $\mtx{B}$ be two matrices of order $n \times n$. The distance between the $i$-th row of $\mtx{A}$ and the $j$-th row of $\mtx{B}$ is denoted and defined as 
$$\ham\left(\mtx{A}({i,*}),\mtx{B}({j,*})\right)=\size{\{k \in [n]:\mat{A}(i,k)\neq \mat{B}(j,k)\}},$$ 
%that is the the number of $k \in [n]$ such that $\mat{A}(i,k)\neq \mat{B}(i,k)$.
 Similarly, $\ham\left(\mtx{A}({*,i}),\mtx{B}({*,j})\right)$ is the distance between the $i$-th column of $\mtx{A}$ and the $j$-th column of $\mtx{B}$. 
 
% between the $i$-th row (column) vector of $\mtx{A}$ and the $j$-th row (column) vector of $\mtx{B}$, and is denoted by $\ham\left(\mtx{A}({i,*}),\mtx{B}({j,*})\right)$ $(\ham\left(\mtx{A}({*,i}),\mtx{B}({*,j})\right))$.
\end{defi}
\begin{obs}[{\bf Expressing $\dist \left(\mat{A},\mat{B}\right)$ as the sum of distance between rows (columns)}]\label{obs:dist}
Let $\mtx{A}$ and $\mtx{B}$ be two $n\times n$  matrices. The matrix distance between $\mtx{A}$ and $\mtx{B}$ is given by 
$$\dist(\mtx{A},\mtx{B})=\sum\limits_{i=1}^n \ham\left(\mtx{A}({i,*}),\mtx{B}({i,*})\right)=\sum\limits_{i=1}^n \ham\left(\mtx{A}({*,i}),\mtx{B}({*,i})\right).$$
\end{obs}

For a given $i\in [n]$, we can approximate $\ham\left(\mtx{A}({i,*}),\mtx{B}({i,*})\right)$ and $\ham\left(\mtx{A}({*,i}),\mtx{B}({*,i})\right)$ using \iporacle{} queries as stated in Lemma~\ref{lem:distrows}. This can be shown by an application of
the well known Johnson-Lindenstrauss Lemma~\cite{DBLP:journals/jcss/Achlioptas03}.
 \begin{lem}[{\bf Estimating the distance between rows of $\mat{A}$ and $\mat{B}$}]\label{lem:distrows}
 Consider \iporacle access to two  $n\times n$ (unknown) matrices $\mtx{A}$ and $
 \mtx{B}$. There is an algorithm $\distrows (i,\alpha,\delta)$, that takes $i\in [n]$ and $\alpha ,\delta\in (0,1)$ as inputs, and reports a $(1\pm \alpha)$-approximation 
 to $\ham\left(\mtx{A}({i,*}),\mtx{B}({i,*})\right)$  with 
 probability at least $1-\delta$, and makes $\Oh\left(\frac{\log n}{\alpha^2}  \log 
 \frac{1}{\delta}\right)$ \iporacle queries to both $\mtx{A}$ and $\mtx{B}$.
 \end{lem}
 
As it is sufficient for our purpose, in the above lemma, we discussed about estimating the distance between rows of $\mat{A}$ and $\mat{B}$ with the same index. However, we note that, a simple modification to the algorithm corresponding to Lemma~\ref{lem:distrows} also works for estimating the distance between any row and/or column pair.

\begin{pre}[{\bf Johnson-Lindenstrauss Lemma}]\label{pre:JL}
Let us consider any pair of points ${\bf u}, {\bf v} \in \R^N$. For a given $\eps \in (0,1)$ and $\delta \in (0,1)$, there is a map $f:\R^N \rightarrow \R^d$ such that $d=\Theta\left(\frac{1}{\eps^2}\log \frac{1}{\delta}\right)$ satisfying the following bound with probability at least $1-\delta$.
\begin{equation}\label{eq:JL}
 (1-\eps)||{\bf u}-{\bf v}||_2^2 \leq  ||f({\bf u})-f({\bf v})||_2^2 \leq (1+\eps)||{\bf u}-{\bf v}||_2^2.
\end{equation}
\end{pre}
\begin{rem}[{\bf An explicit mapping in Johnson-Lindenstrauss Lemma}]\label{rem:map}
An explicit mapping $f:\R^n\rightarrow \R^d$ satisfying Equation~\ref{eq:JL} is as follows. Consider ${\bf r_1},\ldots,{\bf r_d} \in \{-1,1\}^n$ such that each coordinate of every ${\bf r}_i$ is taken from $\{-1,1\}$ uniformly at random. Then for each ${\bf u} \in \{0,1\}^n$, 
$$f({\bf u})=\frac{1}{\sqrt{d}}\left(\dpd{{\bf u}}{{\bf r_1}},\dpd{{\bf u}}{{\bf r_2}},\ldots, \dpd{{\bf u}}{{\bf r_d}}\right).$$
\end{rem}

\paragraph*{Identity testing between two rows.} Now, let us discuss an algorithm where the objective is to decide whether the $i$-th row vectors of matrices $\mat{A}$ and $\mat{B}$ are identical. Observe that $||\mat{A}(i,*)-\mat{B}(j,*)||_2=0$ if and only if $\ham \left(\mat{A}(i,*),\mat{B}(j,*)\right)=0$. Also notice that, for a function $f:\R^n \rightarrow \R^d$ satisfying Equation~\ref{eq:JL}, $||{\bf u}-{\bf v}||_2=0$ if and only if $||f({\bf u})-f({\bf v})||_2=0$. This discussion along with Proposition~\ref{pre:JL} and Remark~\ref{rem:map} imply an algorithm (described inside Observation~\ref{obs:algo-zero}) that can decide whether corresponding rows of $\mat{A}$ and $\mat{B}$ are identical. Observation~\ref{obs:algo-zero} is stated in a more general form than discussed here. Note that the general form will be needed to show Lemma~\ref{lem:distrows}. For this purpose, we define the notion of \emph{projecting a vector in $\{-1,1\}^n$ onto a set $S \subseteq [n]$} as defined below and an observation (Observation~\ref{obs:ip-eval}) about evaluating the projection using an \iporacle query. 
 \begin{defi}[{\bf Vector projected onto a set}]\label{defi:hvec}
 Let $\mtx{A}$ be a  $n\times n$ matrix and $i \in [n]$. For a subset $S\subseteq [n], \mtx{A}(i,*)~\vert_S \in \R^n$ is defined as the vector having $\ell$-th coordinate equals to $\mtx{A}(i,\ell)$ if $\ell \in S$, and $0$, otherwise.
Also consider ${\bf r}\in \{-1,1\}^n$ and a set $S\subseteq [n]$. Then the vector ${\bf r}$ projected onto $S$ is denoted by ${\bf r}\vert_S \in \{-1,0,1\}^n$ and defined as follows: For $\ell \in [n]$, the $\ell$-th coordinate of ${\bf r} \vert_S$ is same as that of ${\bf r}$ if $\ell \in S$, and $0$, otherwise.
 \end{defi}
 
  \begin{obs}\label{obs:ip-eval}
Let $\mtx{A}$ be a  $n\times n$ matrix, $i \in [n]$, ${\bf r} \in \{-1,1\}^n$ and $S \subseteq [n]$. Then $\dpd{\mat{A}(i,*)\vert_S}{{\bf r}}=\dpd{\mat{A}(i,*)}{{\bf r}\vert_S}$. That is, $\dpd{\mat{A}(i,*)}{{\bf r}\vert_S}$ can be evaluated by using a \iporacle query $\dpd{\mat{A}(i,*)}{{\bf r}\vert_S}$ to matrix $\mat{A}$.
 \end{obs}
 
\begin{obs}[Identity testing between rows of $\mat{A}$ and $\mat{B}$]\label{obs:algo-zero}
 Consider \iporacle access to two  $n\times n$ (unknown) matrices $\mtx{A}$ and $\mtx{B}$. There is an algorithm $\ident \left(S,i,\delta\right)$ that takes $i\in [n]$, $S\subseteq [n]$ and $\delta \in (0,1)$ as inputs, and decides whether $ \ham \left(\mat{A}(i,*)~\vert_S,\mat{B}(i,*)~\vert_S\right)=0$ with probability at least $1-\delta$, and makes $\Oh\left(\log \frac{1}{\delta}\right)$ \iporacle queries to both $\mtx{A}$ and $\mtx{B}$.
\end{obs}
\begin{proof}
 Let the vectors ${\mathbf r_1},\ldots,{\mathbf r_d} \in \{-1,1\}^n$ be such that each coordinate of every 
 $\mathbf r_j$, $j=1, \ldots, d$, is taken from $\{-1,1\}$ uniformly at random where $d=\Theta\left(\log \frac{1}{\delta}\right)$. Then the algorithm finds $a_j=\dpd{\mat{A}(i,*)\vert_S}{{\bf r_j}}$ and  $b_j=\dpd{\mat{B}
 (i,*)\vert_S}{{\bf r_j}}$ by making one \iporacle query to each of $\mat{A}$ and $\mat{B}$. This is possible by Observation~\ref{obs:ip-eval}. The algorithm makes $d$ 
  \iporacle queries to each of the matrices $\mat{A}$ and $\mat{B}$. Take ${\bf a}=\frac{1}{\sqrt{d}}(a_1,\ldots,a_d) \in 
 \R^d$ and ${\bf b}=\frac{1}{\sqrt{d}}(b_1,\ldots,b_d) \in \R^d$. By Proposition~
 \ref{pre:JL} and Remark~\ref{rem:map}, $||{\bf a}-{\bf b}||_2=0$  if and only if 
 $||\mat{A}(i,*)~\vert_S,\mat{B}(i,*)~\vert_S||_2=0$. By the definition of distance 
 between a row of one matrix and a row of another matrix (Definition~\ref{defi:dist-row}), note that, $||\mat{A}(i,*)~
 \vert_S-\mat{B}(j,*)~\vert_S||_2=0$ if and only if $\ham \left(\mat{A}(i,*)~
 \vert_S,\mat{B}(j,*)~\vert_S\right)=0$. So, the algorithm finds $||{\bf a} -{\bf 
 b}||_2$ and, reports $||{\bf a}-{\bf b}||_2=0$  if and only if $\ham \left(\mat{A}
 (i,*)~\vert_S,\mat{B}(i,*)~\vert_S\right)=0$. The correctness and query complexity 
 of the algorithm follows from the description itself.% The query complexity of $ \distrows (i,\alpha,\delta)$ is $2d=\Theta\left(\log \frac{1}{\kappa}\right)$.
\end{proof}

\paragraph*{Estimating the distance between rows induced by a set.}
Now, consider the algorithm corresponding to Lemma~\ref{lem:distrows} ($\distrows (\cdot,\cdot,\cdot)$) that can estimate the distance between a row of $\mtx{A}$ and a row of $\mat{B}$. It makes repeated calls to $\ident(\cdot,\cdot,\cdot)$ in a non-trivial way. Also, algorithm $\distrows (\cdot,\cdot,\cdot)$) can be generalized to estimate the distance between a row of $\mat{A}$ and the corresponding row of $\mtx{B}$ projected onto the same set $S \subseteq [n]$, as stated in the following Lemma.
The proof of the following lemma is in Appendix~\ref{sec:append-est-dist}.
 \begin{lem}[{\bf Estimating the distance between rows of $\mat{A}$ and $\mat{B}$ induced by a set $S \subseteq [n]$}]\label{lem:dist-res}
 Consider \iporacle{} access to two  $n\times n$ (unknown) matrices $\mtx{A}$ and $\mtx{B}$. $\resdistrows (S,i,\alpha,\delta)$ algorithm, takes $S \subseteq [n]$, $i \in [n]$ and $\alpha ,\delta\in (0,1)$ as inputs, and reports a $(1\pm \alpha)$-approximation to $\ham\left(\mtx{A}({i,*})~\vert_S,\mtx{B}({i,*})~\vert_S\right)$ with probability at least $1-\delta$, and makes $\Oh\left(\frac{\log n}{\alpha^2}  \log \frac{1}{\delta}\right)$ \iporacle queries to both $\mtx{A}$ and $\mtx{B}$. 
 \end{lem}
 Observe that Lemma~\ref{lem:distrows} is a special case of Lemma~\ref{lem:dist-res} when $S=[n]$. {Algorithm $\distrows (i,\alpha,\delta)$ (corresponding to Lemma~\ref{lem:distrows}) is directly called as a subroutine from $\distsymmmatguess(\mat{A},\mat{B},\vareps,T)$. $\resdistrows (S,i,\alpha,\delta)$ \remove{is not directly called from $\distsymmmatguess(\mat{A},\mat{B},\vareps,T)$, it} is indirectly called from a subroutine to \emph{sample mismatched element almost uniformly} as explained below.}
 
 \paragraph*{Sampling a mismatched element almost uniformly.}
 For a row $i \in [n]$, let $\mbox{{\sc NEQ}}(\mtx{A},\mtx{B},i) = \{j:\mtx{A}(i,j)\neq\mtx{B}(i,j)\}$ denote the set of mismatches.
 Apart from estimating the distance between a row (column) of $\mtx{A}$ and the corresponding row (column) of $\mtx{B}$, we can also sample element from $\mbox{{\sc NEQ}}(\mtx{A},\mtx{B},i)$ \emph{almost uniformly} for any given $i \in [n]$. 
 \begin{defi}[{\bf Almost uniform sample}]\label{defi:almost}
 Let $X$ be a set and $\alpha \in (0,1)$. A $(1\pm\alpha)$-\emph{uniform sample} from $X$ is defined as the sample obtained from a distribution $p$ satisfying $(1-\alpha)\frac{1}{|X|}\leq p(x)\leq (1+\alpha)\frac{1}{|X|}$ for each $x \in X$, where $p(x)$ denotes the probability of getting $x$ as a sample.
 \end{defi}
 \begin{lem}[{\bf Sampling a mismatched element almost uniformly}]\label{lem:almost}
 Consider \iporacle{} access to two  $n\times n$ (unknown) matrices $\mtx{A}$ and $\mtx{B}$. \remove{Also, for $i\in [n]$, $\mbox{{\sc NEQ}}(\mtx{A},\mtx{B},i)$ denotes the set $\{j:\mtx{A}(i,j)\neq\mtx{B}(i,j)\}$.} There exists an algorithm $\randsamp(i,\alpha,\delta)$, described in {Algorithm~\ref{algo:random-samp}}, that takes $i \in [n]$ and $\alpha ,\delta\in (0,1)$ as input, and reports a $(1\pm \alpha)$-uniform sample from the set $\mbox{{\sc NEQ}}(\mtx{A},\mtx{B},i)$ with probability at least $1-\delta$, and makes $\Oh\left(\frac{\log^5 n}{\alpha^2} \log \frac{1}{\delta}\right)$ \iporacle queries to both $\mtx{A}$ and $\mtx{B}$.
 \end{lem}
{ 
Algorithm $\randsamp(\cdot,\cdot,\cdot)$ makes repeated use of $\resdistrows (\cdot,\cdot,\cdot,\cdot)$; the proof of Lemma~\ref{lem:almost} is in Appendix~\ref{sec:append-sample-almost}. We now have all the ingredients -- $\distrows (i,\alpha,\delta)$, $\distsymmmatguess(\mat{A},\mat{B},\vareps,T)$, $\randsamp(i,\alpha,\delta)$ -- to design the final algorithm mentioned in $\distsymmmat(\mat{A},\mat{B},\vareps)$.} 

%\paragraph*{Overview of the proof of Lemma~\ref{lem:matrix-dist-symm}} 
%\noindent 

\paragraph*{Overview of the algorithm.} 
\noindent 
Algorithm $\distsymmmat(\cdot,\cdot,\cdot)$ calls $\distsymmmatguess(\cdot,\cdot,\cdot,\cdot)$ with reduced value of guesses $O(\log n)$ times to bring down the approximation error of matrix distance within limits. 
Algorithm $\distsymmmatguess(\mat{A},\mat{B},\vareps,T)$ discussed in Lemma~\ref{lem:matrix-dist-symm-T} mainly uses subroutines $\distrows (\cdot,\cdot,\cdot)$ and $\randsamp(\cdot,\cdot,\cdot)$ in a nontrivial way. Both of these subroutines use Johnson-Lindenstrauss lemma.

{
Observe that $\distsymmmatguess(\mat{A},\mat{B},\vareps,T)$ estimates $\dist(\mat{A},\mat{B})$ where the approximation gurantee is parameterized by $T$.  By Observation~\ref{obs:dist}, $\dist(\mat{A},\mat{B}) = \sum_{i=1}^n\ham(\mat{A}(i,*),\mat{B}(i,*))$, the sum of the distances among corresponding rows. 
To estimate  $\sum_{i=1}^n\ham(\mat{A}(i,*),\mat{B}(i,*))$,  our algorithm $\distsymmmatguess(\mat{A},\mat{B},\vareps,T)$  considers a partition of the row indices $[n]$ into buckets such that the row indices $i$'s in the same bucket have roughly the same $\ham(\mat{A}(i,*),\mat{B}(i,*))$ values. Now the problem boils down to estimating the sizes of the buckets. To do so, $\distsymmmatguess(\mat{A},\mat{B},\vareps,T)$ finds a random sample $\Gamma$ having $\tOh\left({n}/{\sqrt{T}}\right)$ indices from $[n]$, calls $\distrows (i,\cdot,\cdot)$ for each of the sample in $\Gamma$ and partitions $\Gamma$ into buckets such that $i$'s in the same bucket have roughly the same $\ham(\mat{A}(i,*),\mat{B}(i,*))$ values.  A \emph{large} bucket is one that contains more than a fixed number of row indices. These steps ensure that the sizes of the \emph{large} buckets are approximated well. 
Recall that $\randsamp(i,\alpha,\delta)$ takes $i \in [n]$ and $\alpha ,\delta\in (0,1)$ as input, and reports a $(1\pm \alpha)$-uniform sample from the set $\mbox{{\sc NEQ}}(\mtx{A},\mtx{B},i)$ with probability at least $1-\delta$. {To take care of the \emph{small} buckets, $\distsymmmatguess(\mat{A},\mat{B},\vareps,T)$ calls $\randsamp(i,\cdot,\cdot)$ for \emph{suitable} number of $i$'s chosen uniformly from each large bucket and decides whether the output indices of $\randsamp(i,\cdot,\cdot)$ belong to large or small buckets.} See the pseudocode of our algorithm $\distsymmmatguess(\mat{A},\mat{B},\vareps,T)$ (Algorithm~\ref{algo:random-order} in Appendix~\ref{sec:append-pseudo-codes}) and the following section for the technical details. 
}

\subsection{Proof of Lemma~\ref{lem:matrix-dist-symm-T}}
\label{sec:pf-symm-T}
\noindent
Let us consider the following oracle that gives a probabilistic approximate estimate to the distance between the two corresponding rows of $\mtx{A}$ and $\mtx{B}$; $\mtx{A}$ and $\mtx{B}$ are two unknown $n \times n$ matrices.
\begin{defi}[{\bf Oracle function on the approximate distance between rows}]\label{defi:orc-apx}
Let $\beta,\eta\in (0,1)$. Oracle $\O_{\beta,\eta}$ is a function $\O_{\beta,\eta}:[n]\rightarrow \N$, which when queried with an $i \in [n]$, reports $\O_{\beta, \eta}(i)$. Moreover, 
$$\pr\left(\mbox{for every $i \in [n]$, $\O_{\beta,\eta}(i)$ is a $(1\pm \beta)$-approximation to $\ham\left(\mtx{A}(i,*),\mtx{B}(i,*)\right)$}\right) \geq 1-\eta.$$
\end{defi}

Take $\beta={\eps}/{50}$ and $\eta={1}/{\poly {n}}$ and consider an oracle $\O_{\beta,\eta}$ as defined above. Also, consider a partitioning of the indices in $[n]$ into $t=\Theta\left(\nbuck\right)$ many buckets with respect to $\O_{\beta,\eta}$ such that the $i$'s in the same bucket have {\em roughly} the same $\O_{\beta,\eta}(i)$ values. Let $Y_1,\ldots,Y_t \subseteq [n]$ be the resulting buckets with respect to $\O_{\beta,\eta}$. Formally, for $k\in [t]$, $Y_k=\{i\in [n]:\left(1+\frac{\eps}{50}\right)^{k-1}\leq \O_{\beta,\eta}(i) <\left(1+\frac{\eps}{50}\right)^k\}.$ 
From the definition of $\O_{\beta,\eta}$ and the way we are bucketing the elements of $[n]$, the following observation follows.

\begin{obs}[{\bf Bucketing according to an oracle function}]\label{obs;orc-buck}
Let $\beta={\eps}/{50}$ and $\eta\in (0,1)$. Consider any oracle $\O_{\beta,\eta}:[n]\rightarrow \R$ as defined in Definition~\ref{defi:orc-apx}. Let $Y_1,\ldots,Y_t$ be the buckets with respect to $\O_{\beta,\eta}$. Then
$$\left(1-\frac{\eps}{50}\right)\dist(\mat{A},\mat{B})\leq \sum\limits_{k=1}^t\size{Y_k}\left(1+\frac{\eps}{50}\right)^{k}\leq \left(1+\frac{\eps}{50}\right)^2\dist(\mat{A},\mat{B})$$
holds with probability at least $1-\eta$.
\end{obs} 
\begin{proof}
From Observation~\ref{obs:dist}, $
\dist(\mtx{A},\mtx{B})=\sum_{i=1}^n \ham\left(\mtx{A}({i,*}),\mtx{B}({i,*})\right).$ So, by the definition of $\O_{\beta,\eta}$ along with $\beta={\eps}/{50}$, we have 
\begin{equation}\label{eqn:dist-mat}
 \pr \left(\left(1-\frac{\eps}{50}\right) \dist(\mtx{A},\mtx{B})\leq \sum\limits_{i=1}^n \O_{\beta,\eta}(i) \leq  \left(1+\frac{\eps}{50}\right) \dist(\mtx{A},\mtx{B}) \right) \geq 1-\eta. 
 \end{equation}
 As $Y_1,\ldots,Y_t$ are the buckets with respect to $\O_{\beta,\eta}$, for $k\in [t]$, $Y_k=\{i\in [n]:\left(1+\frac{\eps}{50}\right)^{k-1}\leq \O_{\beta,\eta}(i) <\left(1+\frac{\eps}{50}\right)^k\}.$ So,
 \begin{equation}\label{eqn:dist-orc}
\sum\limits_{i=1}^n \O_{\beta,\eta}(i)  \leq \sum\limits_{k=1}^{t} \size{Y_k}\left(1+\frac{\eps}{50}\right)^k \leq \left(1+\frac{\eps}{50}\right)\sum\limits_{i=1}^n \O_{\beta,\eta}(i)  
 \end{equation}
From Equations~\ref{eqn:dist-mat} and~\ref{eqn:dist-orc}, the following holds with probability at least $1-\eta$.
$$\left(1-\frac{\eps}{50}\right)\dist(\mat{A},\mat{B})\leq \sum\limits_{k=1}^t\size{Y_k}\left(1+\frac{\eps}{50}\right)^{k}\leq \left(1+\frac{\eps}{50}\right)^2\dist(\mat{A},\mat{B}).$$
%Hence, we are done with the proof as $\eps \in \left(0,\frac{1}{2}\right)$.
\end{proof}
The above observation roughly says that $\dist(\mat{A},\mat{B})$ can be estimated if we can approximate $\size{Y_k}$'s.  

\paragraph*{The existence of the oracle.}
Before the description of the algorithm, we note that our algorithm does not need to know the specific 
oracle $\O_{\beta,\eta}:[n]\rightarrow \R$. The existence of some oracle function $\O_{\beta,\eta}:[n]\rightarrow \R$ with respect to which $[n]$ can be partitioned into buckets $Y_1,\ldots,Y_t$ suffices. Our algorithm calls $\distrows(i,\beta,\eta)$ for some $i$'s but for at most once for each $i \in [n]$. Note that $\distrows(i,\beta,\eta)$ is the algorithm (as stated in Lemma~\ref{lem:distrows}) that returns a $\left(1 \pm \beta\right)$-approximation to $\ham{\left(\mat{A}(i,*),\mat{B}(i,*)\right)}$ with probability at least $1-\eta$. So, we can think of $\O_{\beta,\eta}:[n]\rightarrow \R$ such that $\O_{\beta,\eta}(i)=\hat{a_i}$, where $\hat{a_i}$ is the value returned by $\distrows(i,\beta,\eta)$, if the algorithm $\distrows(i,\beta,\eta)$ is called (once). Otherwise, $\O_{\beta,\eta}(i)$ is set to some $(1\pm \beta)$-approximation to $\ham\left(\mat{A}(i,*),\mat{B}(i,*)\right)$~\footnote{This instantiation, for $\O_{\beta,\eta}(i)$'s for which $\distrows(i,\beta,\eta)$'s are never called is to complete the description of function $\O_{\beta,\eta}$. This has no bearing on our algorithm as well as its analysis.}.

\paragraph*{Random sample and bucketing.}
As has been mentioned in the overview of algorithm in Section~\ref{sec:prelim-tech}, the problem of estimating matrix distance boils down to estimating the sizes of the buckets $Y_k$, $k=1, \ldots, t$ and our subsequent action depends on whether the bucket is of \emph{large} or \emph{small} size. But as $\size{Y_k}$'s are unknown, we define a bucket $Y_k$ to be {\em large} or {\em small} depending on the estimate $\size{\hat{Y_k}}$ obtained from a random sample. So, our algorithm starts by taking a random sample $\Gamma \subseteq [n]$ with replacement, where $\size{\Gamma}=\tOh\left({n}/{\sqrt{T}}\right)$, where {$T$ is a guess for $\dist(\mat{A},\mat{B})$}. Now, $\widehat{Y_k}=Y_k\cap \Gamma$, the projection of $Y_k$ on $\Gamma$.

For each $i$ in the random sample $\Gamma$, we call $\distrows(i,\beta,\eta)$ (as stated in 
Lemma~\ref{lem:distrows}) and let $\widehat{a_i}$ be the output. By Lemma~\ref{lem:distrows}, for 
each $i \in \Gamma$, $\widehat{a_i}$ is a $\left(1\pm \frac{\eps}{50}
\right)$-approximation to $\dist\left(\mat{A}(i,*),\mat{B}(i,*)\right)$ with high probability. Based on the values 
of $\widehat{a_i}'s$, we partition the indices in $\Gamma$ into $t$ many 
buckets $\widehat{Y_1},\ldots,\widehat{Y_t}$ such that $i \in \Gamma$ is put into $\widehat{Y_k}$ if and only if $\left(1+\frac{\eps}{50}\right)^{k-1}\leq \widehat{a_i}<\left(1+\frac{\eps}{50}\right)^k$. We define a bucket $Y_k$ to be large or small depending on $\size{\widehat{{Y_k}}} \geq \tau$ or not, where $\tau=\frac{\size{\Gamma}}{n}\frac{\sqrt{\eps T}}{50t}$.

So, if $\size{Y_k}$ is {\em large} (roughly say at least $\sqrt{\eps T}/t$), then it can be well approximated from $\size{\widehat{Y_k}}$. However, it will not be possible to estimate $\size{Y_k}$ from $\size{\widehat{Y_k}}$ if $\size{Y_k}$ is \emph{small}. We explain how to take care of ${Y_k}$'s with {\em small} $\size{Y_k}$.

\remove{
So, if $\size{Y_k}$ is {\em large} (roughly say at least $\sqrt{\eps T}/t$), then it can be well approximated from $\size{\widehat{Y_k}}$, where $\widehat{Y_k}=Y_k\cap \Gamma$, \complain{the projection of $Y_k$ on $\Gamma$}. However, it will not be possible to estimate $\size{Y_k}$ from $\size{\widehat{Y_k}}$ if $\size{Y_k}$ is \emph{small}. We explain how to take care of ${Y_k}$'s with {\em small} $\size{Y_k}$. 

As $\size{Y_k}$'s are unknown, we define a bucket $Y_k$ to be {\em large} or {\em small} depending on the estimate $\size{\hat{Y_k}}$. Let $\tau=\frac{\size{\Gamma}}{n}\frac{\sqrt{\eps T}}{50t}$. We define a bucket $Y_k$ to be {\em large} if $\size{\widehat{{Y_k}}} \geq \tau$, and {\em small}, otherwise. 
}

Let $L\subseteq [t]$ and $S \subseteq [t]$ denote the set of indices for large and small buckets, that is, $L=\{k:Y_k~\mbox{ is large}\}$ and $S=[t]\setminus L$. Also, let $I_L \subseteq [n]$ and $I_S \subseteq [n]$ denote the set of indices of rows present in large and small buckets, respectively. 
From Observation~\ref{obs:dist}, 
$\dist(\mtx{A},\mtx{B})=\sum\limits_{i=1}^n \ham\left(\mtx{A}({i,*}),\mtx{B}({i,*})\right).$
Let us divide the sum $\sum\limits_{i=1}^n \ham\left(\mtx{A}({i,*}),\mtx{B}({i,*})\right)$ into two parts, based on $I_L$ and $I_S$, $d_L$
\begin{eqnarray*}
&&d_L=\sum\limits_{i\in I_L} \ham\left(\mtx{A}({i,*}),\mtx{B}({i,*})\right)=\sum\limits_{k\in L} \sum\limits_{i \in Y_k} \ham\left(\mtx{A}({i,*}),\mtx{B}({i,*})\right)\\
 &\mbox{and}& d_S=\sum\limits_{i\in I_S} \ham\left(\mtx{A}({i,*}),\mtx{B}({i,*})\right)=\sum\limits_{k\in S} \sum\limits_{i \in Y_k} \ham\left(\mtx{A}({i,*}),\mtx{B}({i,*})\right).
\end{eqnarray*}
 That is, $\dist(\mtx{A},\mtx{B})=d_L+d_S$. In what follows, we describe how our algorithm approximates $d_L$ and $d_S$ separately. The formal algorithm $\distsymmmat(\mat{A},\mat{B},\vareps,T)$ is described in Algorithm~\ref{algo:random-order} in Appendix~\ref{sec:append-pseudo-codes}. 

\paragraph*{Approximating $d_L$, the contribution from \emph{large} buckets:}
%Recall that $\distbetrows(i,\beta,\eta)$, as stated in Lemma~\ref{lem:distrows}, that takes $i\in [n]$ and $\beta,\eta \in (0,1)$ as inputs, and reports an $$
We can show in Lemma~\ref{lem:main-inter} (i) and (ii)), for each $k\in L$, $\frac{n}{\size{\Gamma}}\size{\widehat{Y_k}}$ is a $\left(1\pm \frac{\eps}{50}\right)$-approximation to $\size{Y_k}$ with high probability. Recall that $d_L=\sum_{k\in L} \sum_{i \in Y_k} \ham\left(\mtx{A}({i,*}),\mtx{B}({i,*})\right)$, where $L$ denotes the set of indices present in large buckets. Our algorithm $\distsymmmatguess(\mtx{A}, \mtx{B},\vareps,T)$ sets $\widehat{d_L}=\frac{n}{\size{\Gamma}}\sum_{k\in L}\size{\widehat{Y_k}}\left(1+\frac{\eps}{50}\right)^k$ as an estimate for $d_L$. Putting everything together, we show in Lemma~\ref{cl:apx-dl} that the following holds with high probability.%$\widehat{d_L}$ is an $\left(1\pm \frac{\eps}{50}\right)$-approximation to $d_L$ with high probability, that is,
\begin{equation}\label{eqn:largebuck}
\left(1-\frac{\eps}{50}\right) d_L \leq \hat{d_L}\leq \left(1+\frac{\eps}{50}\right)d_L.
\end{equation}

\paragraph*{Approximating $d_S$, the contribution from \emph{small} buckets:} $d_S=\sum_{i\in I_S} \ham\left(\mtx{A}({i,*}),\mtx{B}({i,*})\right)$ can not be approximated directly as in the case of $d_L$. 
To get around the problem of estimating the contribution of small buckets, we partition  $\sum_{i\in I_S} \ham\left(\mtx{A}({i,*}),\mtx{B}({i,*})\right)$ into two parts by projecting row vectors $\mat{A}(i,*)$'s and $\mat{B}(i,*)$'s onto $I_L$ and $I_S$:
 $$d_{SL}=\sum\limits_{i\in I_S} \ham\left(\mat{A}(i,*)\vert_{I_L}, \mat{B}(i,*)\vert_{I_L}\right) \mbox{ and } d_{SS}=\sum\limits_{i\in I_S} \ham\left(\mat{A}(i,*)\vert_{I_S}, \mat{B}(i,*)\vert_{I_S}\right).$$  
 So,  $d_S=d_{SL} + d_{SS}.$ 
 
 As $\mat{A}$ and $\mat{B}$ are symmetric, $d_{SL}=d_{LS} =\sum_{i\in I_L} 
 \ham\left(\mtx{A}({i,*})\vert_{I_S},\mtx{B}({i,*})\vert_{I_S}\right).$ Hence, $d_S=d_{LS}+d_{SS}$. We approximate $d_S$ by arguing that (i) $d_{SS}$ is \emph{small}, and (ii) $d_{LS}$ can be approximated well. 
{Informally speaking, the quantity $d_{SL}$ is all about looking at the large buckets from the small buckets. But as handling small buckets is problematic as opposed to large buckets, we look at the small buckets from the large buckets. Now, as the matrix is symmetric, these two quantities are the same.}
  \begin{description}
 \item[(i) $d_{SS}$ is small:] Observe that $d_{SS}$ can be upper bounded, in terms of $\size{I_S}$, as follows: 
 $$d_{SS}=\sum\limits_{i\in I_S} \ham\left(\mat{A}(i,*)\vert_{I_S}, \mat{B}(i,*)\vert_{I_S}\right)=\size{\{(i,j) \in I_S \times I_S~\mbox{:}~ \mat{A}(i,j)\neq \mat{B}(i,j)\}} \leq \size{I_S}^2.$$ 
 By the definition of $I_S$, it is the set of indices present in small buckets 
($Y_k$'s with $\size{\widehat{Y_k}}\leq \tau$). With high probability, for any 
small bucket $Y_k$, we can show that $\size{Y_k}\leq {\sqrt{\eps T}}/{40t}$. As there are $t$ many buckets, with high probability, $\size{I_S}=\sum_{k\in S}\size{Y_k} \leq \frac{n}{\size{\Gamma}}\tau t \leq 
\frac{\sqrt{\eps T}}{40}$. So, with high probability,
\begin{equation}\label{eqn:small_contr}
d_{SS}\leq \frac{\eps T}{1600}.
\end{equation}
The formal proof of the above equation will be given in Claim~\ref{cl:bound:dss}.
\item[(ii) Approximating $d_{LS}$:]
  For $k \in L$, the set of indices corresponding to large buckets, let $d_{LS}^k$ be the contribution of bucket $Y_k$ to $d_{LS}$, that is, $d_{LS}^k=\sum_{i\in Y_k}
 \ham\left(\mtx{A}({i,*})\vert_{I_S},\mtx{B}({i,*})\vert_{I_S}\right)$. So, $d_{LS}=\sum_{k\in L}d_{LS}^k$, and  
 $d_{LS}$ can be approximated by approximating $d_{LS}^k$ for each $k \in L$. To approximate $d_{LS}^k$, for each $k \in L$, we define $\zeta_k ={d_{LS}^k}/\left({\sum_{i \in Y_k}\ham\left(\mat{A}(i,*),\mat{B}(i,*)\right)}\right)$. We have already argued that $d_{LS}=\sum_{k\in L}d_{LS}^k$ and recall that $\sum_{k \in L}\sum_{i \in Y_k}\ham\left(\mat{A}(i,*),\mat{B}(i,*)\right)=d_L$. So, intuitively, $\zeta_k$ denotes the ratio of the contribution of bucket $Y_k$ to $d_{LS}$ and the contribution of bucket $Y_k$ to $d_L$.
 By our bucketing scheme, for each $i\in Y_k$, $\left(1-\frac{\eps}{50}\right)^{k-1}\leq \ham\left(\mat{A}(i,*),\mat{B}(i,*)\right)\leq\left(1+\frac{\eps}{50}\right)^k$, that is, $\ham\left(\mat{A}(i,*),\mat{B}(i,*)\right)$'s are roughly the same for each $ i \in Y_k$. So, any $d_{LS}^k$ can be  approximated by approximating its corresponding $\zeta_k$. To do so, we express $d_{LS}^k$ combinatorially as follows:  $$d_{LS}^k=\size{\{(i,j):\mat{A}(i,j)\neq \mat{B}(i,j)~\mbox{such that}~i\in Y_k~\mbox{and}~j\in I_S \}}.$$ 
 For each $k \in L$, our algorithm finds a sample $Z_k$ of size $
 \size{\widehat{Y_k}}$ many indices from $\widehat{Y_k}$ with 
 replacement. Then for each $i\in Z_k$, our algorithm calls $
 \randsamp(i,\beta,\eta)$. Recall that  $\randsamp(i,\beta,\eta)$ (as stated in Lemma~\ref{lem:almost}) takes $i \in 
 [n]$ and $\beta,\eta \in (0,1)$ as inputs and returns a $(1\pm \beta)$-uniform sample from the set $ \mbox{{\sc NEQ}}(\mat{A},\mat{B},i)$ with probability at least $1-\eta$. Let $j \in [n]$ be the output of $\mbox{{\sc NEQ}}(\mat{A},\mat{B},i)$.  Then we 
 check whether $j \in I_S$\footnote{The reason for checking $j \in I_S$ can be observed from the definition of $d_{LS}^k=\size{\{(i,j):\mat{A}(i,j)\neq \mat{B}(i,j)~\mbox{such that}~i\in Y_k~\mbox{and}~j\in I_S \}}$.} . Let $C_k$ be the number of elements $i \in Z_k$ whose 
 corresponding call to $\randsamp(i,\beta,\eta)$ returns a $j$ with $j\in I_S$. 
 Our algorithm takes $\hat{\zeta_k}=\frac{C_k}{\size{Y_k}}$ as an estimate for $
 \zeta_k$. We can show that, (in Lemma~\ref{lem:main-inter} (i) and (ii)), for each $k\in L$, $\frac{n}{\size{\Gamma}} \size{\hat{Y_k}}$ is a $\left(1\pm \frac{\eps}{50}\right)$-approximation to $Y_k$. Also, when $T$ is at least a suitable polynomial in $\log n$ and $\frac{1}{\eps}$, we show in Lemma~\ref{lem:main-inter} (iii) and (iv) the followings, respectively:
\begin{itemize}
\item $\zeta_k \geq \frac{\eps}{50}$, then $
 \hat{\zeta_k}$ is a $\left(1\pm \frac{\eps}{40}\right)$-approximation to $
 \zeta_k$ with high probability,
 \item we show that if $\zeta_k \leq \frac{\eps}{50}$, then $\hat{\zeta_k}\leq \frac{\eps}{30}$ holds with high probability.
\end{itemize} 
   Hence, $\hat{d_{LS}}=\frac{n}
 {\size{\Gamma}}\sum\limits_{k \in L}\zeta_k \size{\hat{Y_k}}\left(1+\frac{\eps}{50}\right)^k$ satisfies 
 $\left(1-\frac{\eps}{15}\right)d_{LS}-\frac{\eps}{25}d_L  \leq \hat{d_{LS}} \leq \left(1+\frac{\eps}{15}\right)d_{LS} + \frac{\eps}{25}d_L$
  with high probability. Note that the additive factor in terms of $d_{L}$ is due to the way $\zeta_k$'s are defined.
 \end{description}
 In fact our algorithm $\distsymmmatguess(\mtx{A}, \mtx{B},\vareps,T)$ sets $\hat{d_S}=\hat{d_{LS}}$. The intuition behind setting $\hat{d_S}=\hat{d_{LS}}$ is that $d_{S}=d_{LS}+d_{SS}$ and $d_{SS}$ is small. So, with high probability, 
  \begin{equation}\label{eqn:dsl}
 \left(1-\frac{\eps}{15}\right)d_{LS}-\frac{\eps}{25}d_L  \leq \hat{d_{S}} \leq \left(1+\frac{\eps}{15}\right)d_{LS} + \frac{\eps}{25}d_L.
 \end{equation}
The above will be formally proved in Claim~\ref{cl:bound-dsl}.
By Equations~\ref{eqn:dsl} and~\ref{eqn:small_contr}, we get $\hat{d_S}$ is an estimate for $d_S$ that satisfies the following with high probability.
 \begin{equation}\label{eqn:smallbuck}
\left(1-\frac{\eps}{15}\right)d_{S}-\frac{\eps T}{1600}-\frac{\eps}{25}d_L \leq \hat{d_S} \leq \left(1+\frac{\eps}{15}\right)d_{S}+\frac{\eps}{25}d_{L}
 \end{equation} 
 We will formally show the above equation in Lemma~\ref{cl:apx-ds}.
 \paragraph{Final output returned by our algorithm $\distsymmmatguess(\mtx{A}, \mtx{B},\vareps,T)$:}
  Finally, our algorithm returns $\hat{d}=\hat{d_L}+\hat{d_S}$ as an estimation for $\dist \left(\mat{A},\mat{B}\right)$. Recall that $\dist \left(\mat{A},\mat{B}\right)=d_S+d_L$. From Equations~\ref{eqn:largebuck} and~\ref{eqn:smallbuck}, $\hat{d}$ satisfies, with high probability,
$$\left(1-\frac{\eps}{10}\right)\dist \left(\mat{A},\mat{B}\right)-\frac{\eps}{1600}T \leq \hat{d}\leq \left(1+\frac{\eps}{10}\right)\dist \left(\mat{A},\mat{B}\right).$$  

\paragraph*{The query complexity analysis of our algorithm $\distsymmmatguess(\mtx{A}, \mtx{B},\vareps,T)$: }
Note that the discussed algorithm works when $T$ is at least a suitable polynomial in $\log n$ and 
$1/\eps$. Moreover, the algorithm calls each of $\distrows \left(i,\beta,\eta\right)$ and  $
\randsamp \left(i,\beta,\eta\right)$ for $\tOh\left({n}/{\sqrt{T}}\right)$ times. Note 
that $\beta={\eps}/{50}$ and $\eta={1}/{\poly{n}}$. So, the number of \iporacle queries, made by  
each call to $\distrows \left(i,\beta,\eta\right)$ as well as $\randsamp \left(i,\beta,\eta\right)
$, is $\tOh(1)$ by Lemma~\ref{lem:distrows} and~\ref{lem:almost}. Hence, the number of \iporacle 
queries made by our algorithm is $\tOh\left({n}/{\sqrt{T}}\right)$.\remove{ when $T$ is at least a 
suitable polynomial in $\log n$ \complain{($1/\eps$?)}. Our algorithm ($\distsymmmatguess(\mtx{A}, \mtx{B},\vareps,T)$) handles the otherwise case as follows. It calls $
\distrows \left(i,\beta,\eta\right)$ for each $i \in [n]$ and reports the sum of the outputs as an 
estimate for $\dist{(\mat{A}},{\mat{B})}$. The correctness follows from Observation~\ref{obs:dist} and 
Lemma~\ref{lem:distrows}. The number of queries made by the algorithm is $\tOh(n)=\tOh\left(\frac{n}{\sqrt{T}}\right)$ as we are considering $T$ is at most a suitable polynomial in $\log n$ and $\frac{1}{\eps}$.} The formal proof of the correctness of $\distsymmmatguess(\mtx{A}, \mtx{B},\vareps,T)$ is in Appendix~\ref{sec:append-formal-proof}.

\section{Distance between two arbitrary matrices}\label{sec:arbit}
\noindent
In this Section, we prove our main result (stated as Theorem~\ref{theo:matrix-dist} in Section~\ref{sec:intro}). 
\begin{theo}[Theorem~\ref{theo:matrix-dist} restated]\label{theo:matrix-dist-restate}
There exists an algorithm that has \iporacle query access to unknown matrices $\mtx{A}$ and $\mtx{B}$, takes an $\eps\in (0,1)$ as an input, and returns a $(1\pm \eps)$ approximation to $\dist (\mtx{A},\mtx{B})$ with high probability, and makes $\Oh\left(\frac{n}{\sqrt{\dist (\mtx{A},\mtx{B})}}\mbox{poly}\left(\log n, \frac{1}{\eps}\right)\right)$ queries.
\end{theo}
To prove the above theorem, we use Lemma~\ref{lem:matrix-dist-symm} for estimating $\dist ({\mat{A},\mat{B}})$ when both $\mat{A}$ and $\mat{B}$ are symmetric. Let $\uta$ be a matrix defined as $\uta(i,j)=\mat{A}(i,j)$ if $ i\leq j$, and $\uta(i,j)=\mat{A}(j,i)$, otherwise. Also, let $\lta$ be a matrix defined as $\lta(i,j)=\mat{A}(i,j)$ if $ i \geq j$, and $\lta(i,j)=\mat{A}(j,i)$, otherwise. Similarly, we can also define $\utb$ and $\ltb$ similarly. \remove{Note that {$\mat{A}=\lta+\uta$ and $\mat{B}=\ltb + \utb$}.}  {Observe that $\uta,\lta,\utb$ and $\ltb$ are symmetric matrices, and
$${\dist(\mat{A},\mat{B})=\frac{1}{2}\left[\dist\left(\lta,\ltb \right)+\dist \left(\uta,\utb\right)\right].}$$ }

So, we can report a $(1\pm \vareps)$-approximation to $\dist(\mat{A},\mat{B})$ by finding a $\left(1\pm  \frac{\eps}{2}\right)$-approximation to both $\dist\left(\lta,\ltb \right)$ and $\dist \left(\uta,\utb\right)$ with high probability. This is possible, by Lemma~\ref{lem:matrix-dist-symm}, if we have \iporacle query access to matrices $\lta,\ltb,\uta$ and $\utb$. But we do not have \iporacle query access to $\lta,\ltb,\uta$ and $\utb$ explicitly. However, we can simulate \iporacle query access to matrices $\lta$ and $\uta$ ({$\ltb$ and $\utb$}) with \iporacle query access to matrix $\mat{A}$ ($\mat{B}$), respectively as stated and proved in the observation below. Hence, we are done with the proof of 
Theorem~\ref{theo:matrix-dist-restate}
\begin{obs}
An \iporacle query to matrix $\lta$ ($\uta$) can be answered by using two \iporacle queries to matrix $\mat{A}$. Also, an \iporacle query to $\ltb$ ($\utb$) can be answered by using two \iporacle queries to matrix $\mat{B}$.
\end{obs}
\begin{proof}
We prove how an \iporacle query to matrix $\lta$ can be answered by using two \iporacle queries to matrix $\mat{A}$. Other parts of the statement can be proved similarly.

Consider an \iporacle query $\dpd{\lta(i,*)}{{\bf r}}$ to $\lta$, where $i \in 
[n]$ and ${\bf r}=(r_1,\ldots,r_n) \in \R^n$. Let ${\bf r}^{\leq i}$ and ${\bf r}^{>i}$ in $\R^n$ be two vectors defined as follows: ${r}^{\leq i}_j={ r}_j$ if $j\leq i$, and {${ r}^{\leq i}_j=0$}, otherwise. $r^{>i}_j={r}_j$ if $j>i$, and ${r}^{>i}_j=0$, 
otherwise. Now, we can deduce that
\begin{eqnarray*}
\dpd{\lta(i,*)}{{\bf r}} &=& \sum\limits_{j=1}^{i}\lta(i,j)r_j + \sum\limits_{j=i+1}^{n}\lta(i,j)r_j\\
&=&  \sum\limits_{j=1}^{i}\mat{A}(i,j)r_j + \sum\limits_{j=i+1}^{n}\mat{A}(j,i)r_j
\\
&=& \sum\limits_{j=1}^{n}\mat{A}(i,j)r^{\leq i}_j + \sum\limits_{j=1}^{n}\mat{A}(j,i)r^{>i}_j \\
&=& \dpd{\mat{A}(i,*)}{{\bf r}^{\leq i}} + \dpd{\mat{A}(*,i)}{{\bf r}^{> i}} 
\end{eqnarray*}
From the above expression, it is clear that an \iporacle query of the form $\dpd{\lta(i,*)}{{\bf r}}$ to matrix $\lta$ can be answered by making two \iporacle queries of the form $\dpd{\mat{A}(i,*)}{{\bf r}^{\leq i}}$ and $\dpd{\mat{A}(*,i)}{{\bf r}^{> i}}$ to matrix $\mat{A}$.
%First, we discuss how to get an $\left(1 \pm \frac{\eps}{2}\right)$-approximation to $\dist\left(\lta,\uta \right)$ (by using $\distsymmmat\left(\lta,\ltb,\frac{\vareps}{2}\right)$).

\end{proof}

%we can also find an  $\left(1 \pm \frac{\eps}{2}\right)$-approximation to $\dist\left(\ltb,\utb \right)$
\section{Lower bound results}\label{sec:lower}
\noindent
In this Section, if we ignore polylogarithmic term, we show that (in Theorem~\ref{theo:lb-main}) our algorithm to estimate $\dist{(\mat{A},\mat{B})}$ using \iporacle query is tight. Apart from Theorem~\ref{theo:lb-main}, we also prove that (in Theorem~\ref{theo:lb-decip}) the query complexity of estimating ${\dist(\mat{A},\mat{B})}$ using \iporacledec is quadratically larger than that of using \iporacle{}. The results are formally stated as follows. The lower bounds hold even if the matrices $\mat{A}$ and $\mat{B}$ are symmetric matrices, and one matrix (say $\mat{A}$) is known and one matrix (say $\mat{B}$) is unknown.
% the unknown matrix (say $B$) is known to be a symmetrix matrix. 

\begin{theo}\label{theo:lb-main}
Let $\mat{A}$ and $\mat{B}$ denote the known and unknown (symmetric) matrices, respectively.  Also let $T\in \N$. Any algorithm having \iporacle{} query access to matrix $\mat{B}$, that distinguishes between $\dist{(\mat{A},\mat{B})}=0$ or  $\dist{(\mat{A},\mat{B})}\geq T$ with probability $2/3$, makes $\Omega\left(\frac{n}{\sqrt{T}}\right)$ queries to $\mat{B}$.
\end{theo}

\begin{theo}\label{theo:lb-decip}
Let $\mat{A}$ and $\mat{B}$ denote the known and unknown matrices, respectively.  Also let $T\in \N$. Any algorithm having \iporacledec{} query access to matrix $\mat{B}$, that distinguishes between $\dist{(\mat{A},\mat{B})}=0$ or  $\dist{(\mat{A},\mat{B})}\geq T$ with probability $2/3$, makes $\Omega\left(\frac{n^2}{{T}}\right)$ queries to $\mat{B}$.
\end{theo}
Recall that every \element query to a matrix can be simulated by using a \iporacledec{}. Hence, the following corollary follows.
\begin{coro}\label{coro:lb-me}
Let $\mat{A}$ and $\mat{B}$ denote the known and unknown matrices, respectively.  Also let $T\in \N$. Any algorithm having \element{} query access to matrix $\mat{B}$, that distinguishes between $\dist{(\mat{A},\mat{B})}=0$ or  $\dist{(\mat{A},\mat{B})}\geq T$ with probability $2/3$, makes $\Omega\left(\frac{n^2}{{T}}\right)$ queries to $\mat{B}$.
\end{coro}

\remove{Moreover, $\mat{A}$ is the {\emph null} matrix and $\mat{B}$ is known to be a symmetric matrix.
However, we give a stronger lower bound as stated below when matrix are symmetric and one matrix is known completely apriori. Moreover, the lower bound holds even if the known matrix is the null matrix.}
We prove Theorems~\ref{theo:lb-main} and~\ref{theo:lb-decip} by using a reduction from a problem known as $\disj$ in two party communication complexity.  %The result is formally stated as follows and it will be proved by using a reduction from $\disj$.

%Note that the above lower bound holds even when $\mat{A}$ is an all $1$ matrix and $B$ is known to be symmetrix matrix. 
\paragraph*{Communication complexity}
In two-party communication complexity there are two parties, Alice and Bob, that wish to compute a function $\Pi:\{0,1\}^N \times \{0,1\}^N \to \{0,1\} $. Alice is given ${\bf x} \in \{0,1\}^N$ and Bob is given ${\bf y} \in \{0,1\}^N$. Let $x_i~(y_i)$ denote the $i$-th bit of ${\bf x}~({\bf y})$. While the parties know the function $\Pi$, Alice does not know ${\bf y}$, and similarly, Bob does not know ${\bf x}$. Thus they communicate bits following a pre-decided protocol $\cP$ in order to compute $\Pi({\bf x},{\bf y})$. We say a randomized protocol $\cP$ computes $\Pi$ if for all $({\bf x},{\bf y}) \in \{0,1\}^N \times \{0,1\}^N$ we have $\pr[\cP({\bf x},{\bf y}) = \Pi({\bf x},{\bf y})] \geq 2/3$. The model provides the parties access to common random string of arbitrary length. The cost of the protocol $\cP$ is the maximum number of bits communicated, where maximum is over all inputs $({\bf x},{\bf y}) \in \{0,1\}^N \times \{0,1\}^N$. The communication complexity of the function is the cost of the most efficient protocol computing $\Pi$. For more details on communication complexity, see~\cite{KN97}. We now define $\disj$ function on $N$ bits and state its two-way randomized communication complexity.

\begin{defi}\label{defi:disj}
Let $N \in \N$. The $\disj_N$ on $N$ bits is a function $\disj_N:\{0,1\}^N \times \{0,1\}^N \rightarrow \{0,1\}$ such that $\disj_N({\bf x},{\bf y})=0$ if there exists an $i\in \left[N \right]$ such that $x_i=y_i=1$, and $1$, otherwise.  
\end{defi}
\begin{pre}~\cite{KN97}\label{lem:disj}
The randomized communication complexity of $\disj_N$ is $\Omega(N)$ even if it is promised that there exists at most one $i \in [n]$ such that $x_i=y_i=1$.
\end{pre}
\subsection{Proof of Theorem~\ref{theo:lb-main}}
\noindent
Without loss of generality, assume that $\sqrt{T}$ is an integer that divides $N$. We prove the (stated lower bound) by a reduction from $\disj_N$ where 
$N={n}/{\sqrt{T}}$. Let ${\bf x}$ and ${\bf y}$ in $\{0,1\}^N$ be the inputs of Alice and Bob, respectively. Now consider matrix $\mat{B}$, that depends on both ${\bf x}$ and ${\bf y}$, described as follows.\remove{ Alice and Bob runs the algorithm that has \iporacle access to $\mat{B}$ (as described below) and decides whether $\dist(\mat{A},\mat{B})=0$ or $\dist(\mat{A},\mat{B})=T$ with probability $2/3$.}

\begin{figure}
\begin{center}
\includegraphics[scale=1]{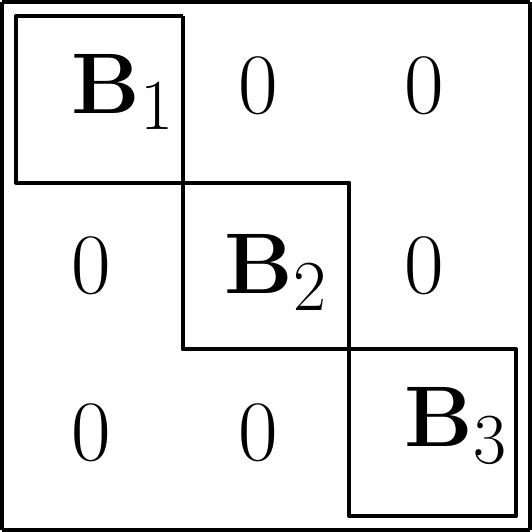}
\caption{A pictorial illustration of a block diagonal matrix $B$ considered in the proof of Theorem~\ref{theo:lb-main}, where $N=3$.}
\label{fig:block}
\end{center}

\end{figure}

\paragraph*{Description of matrices $\mat{A}$ and $\mat{B}$:}
\begin{itemize}
\item[(i)] matrix $\mat{A}$ is the null matrix;
\item[(ii)] matrix $\mat{B}$ is a block diagonal matrix where $\mat{B}_1,\ldots,\mat{B}_N$ are diagonal blocks of order $\sqrt{T} \times \sqrt{T}$ (See {Figure~\ref{fig:block}} for an illustration);
\item[(iii)] Consider $k\in [N]$. If $x_k=y_k=1$, then $\mat{B}_k(i,j)=1$ for each $i,j \in [\sqrt{T}]$, that is, $\mat{B}_k$ is an all-one matrix. Otherwise, $\mat{B}_k$ is a null matrix.
\end{itemize}
%Note that matrix $\mat{B}$ depends on both ${\bf x}$ and ${\bf y}$, and unique with respect to 
 From the description, matrices $\mat{A}$ and $\mat{B}$ are symmetric matrices.  Moreover, if ${\bf x}$ and ${\bf y}$ are disjoint, then all of the $N$ block matrices are null matrices, that is, $\mat{B}$ is also a null matrix. If ${\bf x}$ and ${\bf y}$ are not disjoint, then there is a $k \in \left[N\right]$ such that $\mat{B}_k$ is an all-one matrix, that is, matrix $\mat{B}$ has at least $T$ many $1$s. Recall that here $\mat{A}$ is a null matrix. Hence, $\dist(\mat{A},\mat{B})=0$ if ${\bf x}$ and ${\bf y}$ are disjoint, and $\dist(\mat{A},\mat{B}) \geq T$ if ${\bf x}$ and ${\bf y}$ are not disjoint.
 
 Observe that we will be done with the proof for the stated lower bound by arguing that Alice and Bob can generate the answer to any \iporacle query, to matrix $\mat{B}$, with $2$ bits of communication. Consider a row \iporacle query $\dpd{\mat{B}(i,*)}{{\bf r}}$ to $\mat{B}$ for some $i \in [n]$ and ${\bf r} \in \{0,1\}^n$~\footnote{The proof goes through even if ${\bf r}\in \R^n$.}. From the construction of the matrix $\mat{B}$, there exists a matrix $\mat{B}_j$, for some $j\in [N]$, that completely determines $\mat{B}(i,*)$. Also, observe that, $\mat{B}_j$ depends on $x_j$ and $y_j$ only. So, Alice and Bob can determine 
 $\mat{B}_j$ (hence $\mat{B}(i,*)$) with $2$ bits of communication. As $\mat{B}$ is a symmetric matrix, there is no need to consider column \iporacle queries as such queries can be answered by using row \iporacle queries.

\subsection{Proof of Theorem~\ref{theo:lb-decip}}
\noindent
Here also, we assume that $\sqrt{T}$ is an integer and $\sqrt{T}$ divides $N$, and prove the stated lower bound by a reduction from $\disj_N$ where 
$N={n^2}/{T}$. Let ${\bf x}$ and ${\bf y}$ in $\{0,1\}^N$ be the inputs of Alice and Bob, respectively. Now consider matrix $\mat{B}$, that depends on both ${\bf x}$ and ${\bf y}$, described as follows. In the following description, consider a cannonical mapping $\phi: [N]\rightarrow \left[\frac{n}{\sqrt{T}}\right] \times \left[\frac{n}{\sqrt{T}}\right].$ Note that $\phi$ is known to both Alice and Bob apriori. \remove{ Alice and Bob runs the algorithm that has \iporacledec access to $\mat{B}$ (as described below) and decides whether $\dist(\mat{A},\mat{B})=0$ or $\dist(\mat{A},\mat{B})=T$ with probability $2/3$.}

\begin{figure}
\begin{center}
\includegraphics[scale=1]{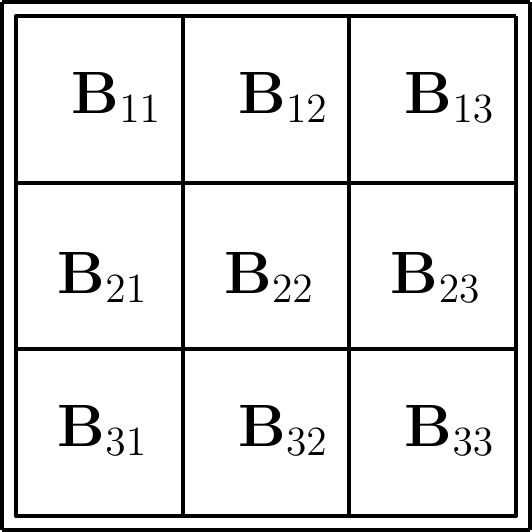}
\caption{A pictorial illustration of a block matrix $\mat{B}$ considered in the proof of Theorem~\ref{theo:lb-decip}, where $N=3$.}
\label{fig:block1}
\end{center}

\end{figure}

\paragraph*{Description of matrices $\mat{A}$ and $\mat{B}$:}
\begin{itemize}
\item[(i)] matrix $\mat{A}$ is an all $1$ matrix;
\item[(ii)] matrix $\mat{B}$ is a block matrix where $\mat{B}_{ij}$'s $\left(i,j \in \left[\frac{n}{\sqrt{T}}\right]\right)$ are  blocks of order $\sqrt{T} \times \sqrt{T}$ (See {Figure~\ref{fig:block1}} for an illustration);
\item[(ii)] Consider $k \in [N]$. If $x_k=y_k=1$, then $\mat{B}_{\phi(k)}=0$~\footnote{Here we abuse the 
notation slightly. If $\phi(k)=(i,j)$, we denote $\mat{B}_{ij}$ by $\mat{B}_{\phi(k)}.$} for each $i,j \in 
[\sqrt{T}]$, that is, $\mat{B}_{\phi{(k)}}$ is an all $0$ matrix. Otherwise, $\mat{B}_{\phi (k)}$ is an all $1$ 
matrix.
\end{itemize}
%Note that matrix $\mat{B}$ depends on both ${\bf x}$ and ${\bf y}$, and unique with respect to 
 From the description, matrices $\mat{A}$ and $\mat{B}$ are symmetric matrices.  Moreover, if ${\bf x}$ and ${\bf y}$ are disjoint, then all of the $N$ block matrices are all $1$ matrices, that is, $\mat{B}$ is also an all $1$ matrix. If ${\bf x}$ and ${\bf y}$ are not disjoint, then there is (exactly) one $k \in \left[N\right]$ such that $\mat{B}_k$ is a null matrix, that is, matrix $\mat{B}$ has exactly $T$ many $0$s. Recall that here $\mat{A}$ is a null matrix. Hence, $\dist(\mat{A},\mat{B})=0$ if ${\bf x}$ and ${\bf y}$ are disjoint, and, $\dist(\mat{A},\mat{B}) = T$ if ${\bf x}$ and ${\bf y}$ are not disjoint.
 
 Observe that we will be done with the proof for the stated lower bound by arguing that Alice and Bob can generate the answer to any \iporacledec query, to matrix $\mat{B}$, with $2$ bits of communication. Consider a  row \iporacledec query $\dpd{\mat{B}(i,*)}{{\bf r}}$ to $\mat{B}$ for some $i \in [n]$ and ${\bf r} \in \{0,1\}^n$. Without loss of generality, assume that ${\bf r}$ is not a null matrix, as in this case Alice and Bob can decide $\dpd{\mat{B}(i,*)}{{\bf r}}=0$ trivially without any communication. Consider a partition of ${\bf r}$ into ${N}/{\sqrt{T}}$ subvectors ${\bf r}_1,\ldots, {\bf r}_{n/\sqrt{T}}\in \{0,1\}^{\sqrt{T}}$ in the natural way (See {Figure~\ref{fig:block1}} for an illustration). \remove{Observe that 
$$\dpd{\mat{B}(i,*)}{{\bf r}}=\sum\limits_{j \in \left[\frac{n}{\sqrt{T}}\right]} \dpd{\mat{B}(i,*)}{{\bf r}_j}.$$ } Now we analyze by making two cases. If two of the ${\bf r}_j$s are not null vectors, from the construction of the matrix $\mat{B}$, then $\dpd{\mat{B}(i,*)}{{\bf r}} >0$. So, in this case, Alice and Bob report  $\dpd{\mat{B}(i,*)}{{\bf r}} \neq 0$ without any communication between them. Now consider the case when exactly one of the ${{\bf r}_j}$ is not a null vector. Once again, from the construction of $\mat{B}$, {deciding whether $\dpd{\mat{B}(i,*)}{{\bf r}}=0$ is equivalent to deciding whether a particular block (say $\mat{B}_{xy}$) is a null matrix. It is because ${\bf r}$ is not a null vector, and each block in $\mat{B}$ is either a null matrix or an all $1$ matrix}. Let $k \in \left[\frac{n}{\sqrt{T}}\right]$ be such that $\phi(k)=(x,y)$.  Note that $\mat{B}_{xy}$ is a null matrix if and only if $x_k=y_k=1$. So, Alice and Bob can determine whether $\dpd{\mat{B}(i,*)}{{\bf r}} =0$ with $2$ bits of communication. As $\mat{B}$ is a symmetric matrix, there is no need to consider column \iporacledec queries as such queries can be answered by using row \iporacle queries.

\remove{
there exists at most one $j \in \left[\frac{N}{\sqrt{T}}\right]$ such that each ${\bf r}_{i}$ $\left(i \in \left[\frac{n}{\sqrt{T}}\right]\right)$ has exactly one row in common with $B_{ij}$. Once there exists at most one matrix $\mat{B}_j$, for some $j\in [N]$, that is null matrix. Also, observe that, $\mat{B}_j$ depends on $x_j$ and $y_j$ only. 
 $\mat{B}_j$ (hence $\mat{B}(i,*)$) with $2$ bits of communication. 

}

%\input{appln.tex}
%\appendix

\section{Conclusion}
\label{sec:conclude}
\noindent
Recall that in an \iporacle as well as in \iporacledec queries, a vector ${\bf v} \in \{0,1\}^n$ is given as input along with an index for row or column of an unknown matrix.
Let $\mbox{\iporacle}_\R$ and $\mbox{\iporacledec}_\R$ be the extension of \iporacle and \iporacledec when ${\bf v}$ is a vector in $\R^n$. Now let us have a look into Table~\ref{table:results-conflict-and-est-graph}. The lower bound of $\Omega(n/\sqrt{D})$, on the number of \iporacle queries to estimate $D=\dist(\mat{A},\mat{B})$, also holds even when we have an access to $\mbox{\iporacle}_\R$ oracle. This also implies a lower bound of $\Omega(n/\sqrt{D})$ on the number of \iporacledec queries to solve the problem at hand. But our lower bound proof of $\Omega(n^2/D)$ on the number of \iporacledec queries does not work when we have access to $\mbox{\iporacle}_\R$ oracle. So, we leave the following problem as open.
\paragraph*{Open problem:} What is the query complexity of estimating $\dist(\mat{A},\mat{B})$ when we have $\mbox{\iporacledec}_\R$ access to matrices $\mat{A}$ and $\mat{B}$?

\newpage
\bibliographystyle{alpha}
\bibliography{reference}
\newpage

\appendix

%%%%%%%%%%% Pseudo-codes %%%%%%%%%%%%%%
\section{Pseudocode for $\distsymmmatguess(\mat{A},\mat{B},\vareps,T)$}
\noindent
The pseudocode for $\distsymmmatguess(\mat{A},\mat{B},\vareps,T)$ (Algorithm for distance between two (unknown) $n \times n$ symmetric matrices $\mtx{A}$ and $\mtx{B}$) is presented in Algorithm~\ref{algo:random-order}. The pseudocode is presented w.r.t. the following description.
 \begin{itemize}
 \item $\Gamma\subseteq [n]$ denotes a set of random samples;
 \item $t=\Theta\left( \log _{\eps/50} n\right)$, $\tau= \frac{\size{\Gamma}}{n}\frac{\sqrt{\vareps T}}{40 t}$, $\beta=\frac{\eps}{50}$ and $\eta=\frac{1}{\poly{n}}$;
 \item $\distrows(i,\beta,\eta)$ is the algorithm, as stated in Lemma~\ref{lem:distrows}, that takes $i\in [n]$ and $\beta,\eta \in (0,1)$ as inputs and returns a $(1 \pm \beta)$-approximation to $\ham \left(\mat{A}(i,*),\mat{B}(i,*)\right)$ with probability $1-\eta$; moreover, our algorithm calls $\distrows(i,\beta,\eta)$ for a particular $i\in [n]$ at most once;
 \item $\randsamp(i,\beta,\eta)$ is the algorithm, as stated in Lemma~\ref{lem:almost}, that takes $i\in [n]$ and $\beta,\eta \in (0,1)$ as inputs and returns a $(1\pm \beta)$-uniform sample from $\mbox{{\sc NEQ}}(\mat{A},\mat{B},i)=\{j:\mat{A}(i,j)\neq \mat{B}(i,j)\}$ with probability at least $1-\eta$;
 \item Oracle $\O_{\beta,\eta}:[n]\rightarrow \R$ is a function, as defined in Definition~\ref{defi:orc-apx}, such that $\O_{\beta,\eta}(i)$ equals to $\hat{a_i}$, the value returned by $\distrows(i,\beta,\eta)$, if the algorithm $\distrows(i,\beta,\eta)$ is called (once). Otherwise, $\O_{\beta,\eta}(i)$ is set to some $(1\pm \beta)$-approximation to $\ham\left(\mat{A}(i,*),\mat{B}(i,*)\right)$.
 \item Let $Y_1,\ldots,Y_t \subseteq [n]$ be the buckets into which $[n]$ is partitioned, where $$Y_k=\{i\in [n]: \left(1+\frac{\eps}{50}\right)^{k-1}\leq \O_{\beta,\eta}(i)< \left(1+\frac{\eps}{50}\right)^{k}\}.$$
 \end{itemize}
\label{sec:append-pseudo-codes}
\begin{algorithm}[H]

\caption{$\distsymmmatguess(\mat{A},\mat{B},\vareps,T)$: Algorithm for distance between two (unknown)  $n\times n$ symmetric matrices $\mtx{A}$ and $\mtx{B}$.}
\label{algo:random-order}
\KwIn{\iporacle{} access to unknown  matrices $\mtx{A}$ and $\mtx{B}$, parameters $T$ and $\vareps$.}
\KwOut{$\widehat{d}$, that is, a $(1\pm \vareps)$ approximation to ${\dist(\mtx{A},\mtx{B})}$.}

 \remove{\If{($T$ is less than that of suitable polynomial in $\log n$ and $1/\eps$)}
 {
For each $i \in [n]$, call $\distrows(i,\beta,\eta)$ and $\hat{a_i}$ be the output. 

Report $\sum_{i=1}^n \hat{a_i}$ as the output.
 }}
%\Else
{
 Sample a set $\Gamma\subseteq [n]$ of $\tOh\left(\frac{n}{\sqrt{T}}\right)$ indices of rows with replacement.

 \For{(Each $i\in \Gamma$)}{ 
 $$\mbox{Call $\distrows(i,\beta,\eta)$, and let $\widehat{a_i}$ be the output.}$$}
 %// $\widehat{a_i}$ is an $\left(1\pm \frac{\vareps}{10}\right)$ to $\ham(\mtx{A}(i,*),\mtx{B}(i,*))$ with probability at least $1-\eta$.
 
 \For{(Each $k\in [t]$)}
 {
 $$\mbox{Determine $\hat{Y_k}$ defined as $\hat{Y_k}=\{i\in \Gamma:\left(1+\frac{\eps}{50}\right)^{k-1}\leq \widehat{a_i} < \left(1+\frac{\eps}{50}\right)^k\}.$}$$}
 
 Determine the set of large and small buckets $L$ and $S$, respectively, defined as follows:
 $$ \mbox{$L=\{k\in [t]:\size{\hat{Y_k}}\geq \tau \}$ and  $S=[t]\setminus L$.}$$
 
$\mbox{ Set $\hat{d_L}=\frac{n}{\size{\Gamma}}\sum\limits_{k\in L}\size{\hat{Y_k}}\left(1+\frac{\eps}{50}\right)^k$.}$
 
 \For{(Each $k\in L$)}
 {
 Initialize a counter $C_k=0$.
 
 Generate $Z_k$ of $\size{\hat{Y_k}}$ many random samples with replacement from $\hat{Y_k}$. 
  
 \For{(Each $i\in Z_k$)}
 { 
  $$\mbox{Call $\randsamp(i,\beta,\eta)$. Let $j_i \in [n]$ be the output}.$$
  If $j_i\in \Gamma$, that is if $\distrows(j_i,\beta,\eta)$ has been called previously, the value of $\widehat{a_{j_i}}$ is already known. Otherwise, call $\distrows(j_i,\beta,\eta)$ and set the output to be $\widehat{a_{j_i}}$. 
  
  // This step ensures that $\distrows(\ell,\beta,\eta)$ is called at most once for a particular $\ell \in [n]$.
  
  Increment $C_k$ by $1$ if and only if $\hat{a_{j_i}}<\tau$, that is, $j_i \in I_S$.
  }
  Set $\widehat{\zeta_k}=\frac{C_k}{\size{\hat{Y_k}}}$.
 
 }
  $\mbox{Set $\hat{d_S}=\frac{n}{\size{\Gamma}}\sum\limits_{k\in L}\hat{\zeta_k}\size{\hat{Y_k}}\left(1+\frac{\eps}{50}\right)^k$.}$
 
 $\mbox{Return $\hat{d_L}+\hat{d_S}$ as the output $\widehat{d}$.}$
 }
\end{algorithm}

%Proof of Lemma on sampling a mismatched element almost uniformly 

\section{Proof of Lemma~\ref{lem:dist-res}}
\label{sec:append-est-dist}
\noindent
 \begin{lem}[Lemma~\ref{lem:dist-res} restated]
 Consider \iporacle{} access to two  $n\times n$ (unknown) matrices $\mtx{A}$ and $\mtx{B}$. There exists an algorithm $\resdistrows (S,i,\alpha,\delta)$ that takes $S \subseteq [n]$, $i \in [n]$ and $\alpha ,\delta\in (0,1)$ as inputs, and reports a $(1\pm \alpha)$-approximation to $\ham\left(\mtx{A}({i,*})~\vert_S,\mtx{B}({i,*})~\vert_S\right)$ with probability at least $1-\delta$, and makes $\Oh\left(\frac{1}{\alpha^2} \log n \log \frac{1}{\delta}\right)$ \iporacle queries to both $\mtx{A}$ and $\mtx{B}$. 
 \end{lem}
 As noted in Section~\ref{sec:prelim-tech}, $\distrows (i,\alpha,\delta)$ makes repeated calls to $\ident \left(S,i,\kappa\right)$ (the algorithm corresponding to Observation~\ref{obs:algo-zero}). We will prove Lemma~\ref{lem:dist-res} by using a result on \emph{estimating the size of an unknown subset} of a known universe using \emph{emptiness query oracle} access defined below. {We will show that \iporacle can simulate this oracle, and thus can solve the subset estimation problem.} 
 \paragraph*{Emptiness Query Oracle and subset size estimation.} Let us first define the emptiness query oracle. 
 \begin{defi}
 Let $U$ be a known universe and $X\subseteq U$ be unknown. Emptiness Query Oracle takes a query set $Q \subseteq U$ as input and lets us know whether $Q \cap X=\emptyset$.
 \end{defi}
 {The following proposition mentions the query complexity of subset size estimation using emptiness queries.}
 \begin{pre}[Estimation of size of an unknown subset using emptyness queries (Lemma~2.9 in ~\cite{DBLP:journals/corr/abs-1711-07567})]\label{pre:emptyness}
 Let $U$ be a universe of size $n$. There exists an algorithm that has an emptiness query oracle access to an unknown subset $X \subseteq U$, takes parameters $\alpha,\delta \in (0, 1)$,  returns a $(1 \pm \alpha)$-approximation to $\size{X}$ with probability at least $1-\delta$, and makes $\Oh\left(\frac{\log n}{\alpha^2} \log \frac{1}{\delta}\right)$ emptiness queries.
 \end{pre}
 {Now, we come back to the proof of Lemma~\ref{lem:dist-res} by simulating the emptiness query oracle by \iporacle{}.}
 \begin{proof}[Proof of Lemma~\ref{lem:dist-res}]
We will set up the framework of Proposition~\ref{pre:emptyness}. Let 
$U=[n]$ and $X\subseteq [n]$ is the unknown set defined as, $X=\{k\in S:\mat{A}(i,k)\neq \mat{B}(i,k)\}$. Observe that 
$ \ham \left(\mat{A}(i,*)~\vert_S,\mat{B}(i,*)~\vert_S\right)=\size{X}.$ By Proposition~\ref{pre:emptyness}, we can estimate $\size{X}= \ham \left(\mat{A}(i,*)~\vert_S,\mat{B}(i,*)~\vert_S\right)$ by using $\Oh\left(\frac{\log n}{\alpha^2} \log \frac{1}{\delta}\right)$ emptiness queries with probability at least $1-\frac{\delta}{2}$. Hence, we will be done by showing how to answer all the emptiness queries by using \iporacle{} queries with probability $1-\frac{\delta}{2}$. Let $Q \subseteq [n]$ be (an emptiness query) such that we want to know whether $Q \cap X=\emptyset$. Observe that $Q \cap X=\emptyset$ if and only if $ \ham \left(\mat{A}(i,*)~\vert_Q,\mat{B}(i,*)~\vert_Q\right)=0$. Also,  we can know whether $ \ham \left(\mat{A}(i,*)~\vert_Q,\mat{B}(i,*)~\vert_Q\right)=0$ or not with probability $1-\frac{\delta}{2n}$ by calling $\ident \left(S,i,\delta/2n\right)$ (the algorithm corresponding to Observation~\ref{obs:algo-zero}) and making $\Oh\left(\log \frac{n}{\delta}\right)$ many \iporacle{} queries. Hence, each emptiness query can be simulated by using $\Oh\left(\log \frac{n}{\delta}\right)$ many \iporacle{} queries with probability $1-\frac{\delta}{2n}$. As there is no need to make more than $\Oh(n)$ many emptiness queries (irrespective of input parameters $\alpha$ and $\delta$), we get correct answers to all the emptiness queries with probability at least $1-\frac{\delta}{2}$.

As we are making $\Oh\left(\frac{\log n}{\alpha^2} \log \frac{1}{\delta}\right)$ many emptiness queries and each emptiness query is simulated by using $\Oh\left(\log \frac{n}{\delta}\right)$ many \iporacle{} queries, the stated query compplexity for $\resdistrows (S,i,\alpha,\delta)$ follows.
\end{proof}

\section{Proof of Lemma~\ref{lem:almost}}
\label{sec:append-sample-almost}
\noindent  
To prove Lemma~\ref{lem:almost}, we need notions of \emph{hierarchical subsets} of $[n]$  defined in Definition~\ref{defi:hset} and an observation (Observation~\ref{obs:hset}) about hierarchical subsets.

  \begin{lem}[Lemma~\ref{lem:almost} restated]
 Consider \iporacle{} access to two  $n\times n$ (unknown) matrices $\mtx{A}$ and $\mtx{B}$. \remove{Also, for $i\in [n]$, $\mbox{{\sc NEQ}}(\mtx{A},\mtx{B},i)$ denotes the set $\{j:\mtx{A}(i,j)\neq\mtx{B}(i,j)\}$.} There exists an algorithm $\randsamp(i,\alpha,\delta)$, described in Algorithm~\ref{algo:random-samp}, that takes $i \in [n]$ and $\alpha ,\delta\in (0,1)$ as input, reports a $(1\pm \alpha)$-uniform sample from the set $\mbox{{\sc NEQ}}(\mtx{A},\mtx{B},i)$ with probability at least $1-\delta$, and makes $\Oh\left(\frac{1}{\alpha^2} \log \frac{n}{\delta}\right)$ \iporacle queries to both $\mtx{A}$ and $\mtx{B}$.
 \end{lem}

\begin{defi}[{\bf Hierarchical subsets}]\label{defi:hset}
For $j\in [\log n]\cup \{0\}$, the $k$-th with $k\in \left[2^j\right]$ hierarchical subset of label $j$ is denoted and defined as $S_{j}^k=\{(k-1)\frac{n}{2^j}+1,\ldots,k\frac{n}{2^j}\}$. The set of \emph{hierarchical} subsets of $[n]$ with label $j$ is denoted and defined as $\cH_j=\{S_{j,k}:k\in \left[2^j\right]\}$. The set of hierarchical subsets of $[n]$ is denoted an defined as by $\cH= \bigcup\limits_{j=0}^{\log n}\cH_j=\{S_{j}^k:j\in [\log n]\cup \{0\}~\mbox{and}~k\in \left[2^j\right]\}$.  
 \end{defi}
 \begin{obs}\label{obs:hset}
Consider the set of hierarchical subsets $\cH$ of $[n]$ defined as above. Then
\begin{itemize}
\item[(i)] $\size{\cH}=2n-1$;
\item[(ii)] For any $j\in [\log n]\cup \{0\}$, the set of hierarchical subsets of label $j$ forms a partition of $[n]$;
\item[(iii)] For every $j \in [\log n-1]\cup \{0\}$ and $k \in [2^j]$, the $k$-th hierarchical subset in label $j$ is the disjoint union of two hierarchical subsets, that is, $(2k-1)$-th and $2k$-th hierarchical subset of label $j$. In other words,
$S_{j}^k=S_{j+1}^{2k-1} \sqcup S_{j+1}^{2k}$; 
\item[(iv)] For any $\ell \in [n]$ and for any $j\in [\log n]\cup \{0\}$, there is exactly one $k_j \in [2^j]$ such that $\ell$ is present in the $k_j$-th hierarchical subset of label $j$, that is, in $S_{j}^{k_j}$. Let $I(j,\ell)\in [2^j]$ be such that $I(j,\ell)$-th hierarchical subset of label $j$ has $\ell$ in it. 
\item[(v)] $I({\log n},\ell)=\ell$;
\item[(vi)]  For any $j\in [\log n-1]\cup \{0\}$, $I(j,\ell)=2I(j-1,\ell)-1$ or $I(j,\ell)=2I(j-1,\ell)$. That is, $I(j-1,\ell)=\lceil \frac{I(j,\ell)}{2} \rceil $.

%If $\ell$ in present in $k_j$-th and $k_{j+1}$-th subset of label $j$ and label $j+1$, respectively, then $k_{j+1}=2k_j$ or $k_{j+1}=2k_j-1$, that is, $k_j=\lceil \frac{k_{j+1}}{2}\rceil$.
\end{itemize}
 \end{obs}
 Apart from Definition~\ref{defi:hset} and Observation~\ref{obs:hset}, we also need Lemma~\ref{lem:dist-res} that says that the distance between a row vector of $\mat{A}$ projected onto a set $S \subseteq [n]$ with the corresponding row vector of $\mat{B}$ projected onto $S$, can be approximated. 
 Now, we have all intermediate stuff to show Lemma~\ref{lem:almost}. %For a set $S$ in the set $\cH$ of hierarchical subsets, let us define $\mbox{{\sc NEQ}}(\mat{A},\mat{B},S,i)=\mbox{{\sc NEQ}}(\mat{A},\mat{B},i)\cap S$
 \begin{algorithm}
\caption{$\randsamp \left(i,\alpha,\delta \right)$: Algorithm for sampling a $(1 \pm \alpha )$-approximation sample from $\mbox{{\sc NEQ}}(\mtx{A},\mtx{B},i)$.}
\label{algo:random-samp}
\KwIn{\iporacle access to unknown matrices $\mtx{A}$ and $\mtx{B}$, parameters $i\in [n]$ and $\alpha \in (0,1)$.}
\KwOut{A $(1 \pm \alpha)$-approximation sample from $\mbox{{\sc NEQ}}(\mtx{A},\mtx{B},i)$.}
Set $\alpha'=\frac{\alpha}{50 \log n}$ and $\delta'=\frac{\delta}{50 \log n}$.
%Call $\resdistrows \left(S_{0}^1,i,\alpha',\delta'\right)$ and let $d_0^1$ be the output.//$S_{0}^{1}=[n]$ is the first and only set having label $0$ in the set $\cH$ of hierarchical subsets of $[n]$.

Set $k_{0}=1$

\For{($j=1$ to $\log n$)}
{
Call $\resdistrows \left(S_{j}^{2k_{j-1}-1},i,\alpha',\delta'\right)$ and $\resdistrows \left(S_{j-1}^{2k_{j-1}},i,\alpha',\delta',\right)$. Let $d_1^j$ and $d_2^j$ be the respective outputs.

{// $\resdistrows \left(\cdot, \cdot, \cdot, \cdot \right)$ is the algorithm as stated in Lemma~\ref{lem:dist-res}.}

// Note that $S_{j}^{2k_{j-1}-1}$ and $S_{j}^{2k_{j-1}}$ are $(2k_{j-1}-1)$-th and $2k_j$-th hierarchical subsets of $[n]$ (See Definition~\ref{defi:hset}). 

%Let $d_{j-1}^{k_{j-1}}$ be the output of the call to $\resdistrows \left(S_{j-1}^{k_{j-1}},i,\alpha',\delta'\right)$ in the previous iteration.
 Set $k_j=2k_{j-1}-1$ with probability $\frac{d_1^j}{d_1^j+d_2^j}$ and $k_j$ with the remaining probability.
}
Report $k_{\log n}$ as the output.
\end{algorithm}
\begin{proof}[Proof of Lemma~\ref{lem:almost}]
Algorithm $\randsamp(i,\alpha,\delta)$  sets $\alpha'=\frac{\alpha}{10 \log n}$ and $\delta'=\frac{\delta}{50 \log n}$ and goes over $\log n$ many iterations. In the first iteration, it calls for $\resdistrows \left(S_{1}^{1},i,\alpha',\delta'\right)$ and $
\resdistrows \left(S_{1}^{2},\alpha',\delta',i\right)$. Let $d_{1}^{1}$ and $d_{1}
^{2}$ be the respective outputs. Then we select $S_1^1$ with probability $
\frac{d_1^1}{d_1^2+d_2^1}$ and on $S_1^2$ with the remaining probability. The algorithm sets $k_1=1$ or $2$ (as an indicator of which subset out of $S_1^1$ and $S_1^2$ is selected) depending on whether $S_1^1$ or $S_1^2$ is selected, and recurse on the selected subset in the second iteration. In the $j$-th iteration, note that, the algorithm calls $\resdistrows \left(S_{j}^{2k_{j-1}-1},i,\alpha',\delta'\right)$ and $\resdistrows \left(S_{j}^{2k_{j-1}},i,\alpha',\delta',\right)$. Let $d_1^j$ and $d_2^j$ be the respective outputs. Note that $S_{j}^{2k_{j-1}-1}$ and $S_{j}^{2k_{j-1}}$ are $(2k_{j-1}-1)$-th and $2k_j$-th hierarchical subsets of $[n]$ (See Definition~\ref{defi:hset}). The algorithm selects $S_j^{2k_{j-1}-1}$ with probability $\frac{d_1^j}{d_1^j+d_2^j}$ and $S_{j}^{2k_j}$ with the remaining probability. Then the algorithm sets $k_j=2k_{j-1}-1$ and $k_j=2k_{j-1}$ as an indicator of which set out of 
$S_{j}^{2k_j-1}$ and $S_j^{2k_j}$ is selected.
After executing all the $\log n$ iterations, $\randsamp \left(i,\alpha,\delta \right)$ reports $k_{\log n}$ as the output. A pseudocode for $\randsamp(i,\alpha,\delta)$ is given in Algorithm~\ref{algo:random-samp}.

 $\randsamp(i,\alpha,\delta)$ calls for $\resdistrows(S,i,
\alpha',\delta')$ $2\log n-1$ times in total by taking different parameters in 
place of $S$. By Lemma~\ref{lem:dist-res}, we make $\Oh \left(\frac{1}{(\alpha')^2} \log \frac{1}\delta'\right)$ many \iporacle queries to matrices $\mat{A}$ and $\mat{B}$. So, the total number of \iporacle queries made by $\randsamp(i,\alpha,\delta)$ is $\Oh \left(\frac{\log ^4 n}{\alpha^2}\log \frac{n}{\delta}\right)$. Let $\cE$ be the event that all the calls to $\resdistrows$ gives 
desired output. Observe that $\Pr(\cE) \geq 1-\delta'(2\log n-1) \geq 1-\delta$. 
We show the correctness proof for $\randsamp(i,\alpha,\delta)$ on the conditional space that $\cE$ holds. Particularly, we will show that the algorithm returns an $(1 \pm \alpha)$-
approximate sample from set $\mbox{{\sc NEQ}}(\mat{A},\mat{B},i)$ when $\cE$ holds. From the description of Algorithm~\ref{algo:random-samp}, $k_{\log n}$ is the output which is a random variable. In what follows, we will show that $k_{\log n}$ is a $\left(1 \pm \alpha \right)$-approximate sample from $\mbox{{\sc NEQ}}(\mat{A},\mat{B},i)$, that is, for any $\ell \in \mbox{{\sc NEQ}(\mat{A},\mat{B})}$,

\begin{equation}
 (1-\alpha)\frac{1}{\size{\mbox{{\sc NEQ}(\mat{A},\mat{B},i)}}} \leq \pr(k_{\log n}=\ell) \leq (1+\alpha)\frac{1}{\size{\mbox{{\sc NEQ}(\mat{A},\mat{B},i)}}}.
 \end{equation}

From Observation~\ref{obs:hset} (iv), recall that $I(j,\ell)\in [2^j]$ be such that such that $I(j,\ell)$-th hierarchical subset of label $j$ has $\ell$ in it. Moreover, $I(\log n,\ell)=\ell$ by Observation~\ref{obs:hset} (v). So, we approximate the probability of the event $k_{\log n}=I(\log n,\ell)$ instead of $k_{\log n}=\ell$. Consider the following observation that follows from the description of our algorithm along with Observation~\ref{obs:hset}.

\begin{obs}\label{obs:sampled-all}
\begin{itemize}
\item[(i)] $k_{\log n}=\ell$ if and only if $k_{j}=I(j,\ell)~\forall ~j \in [\log n]\cup \{0\}$;
\item[(ii)] In general, for any $j\in [\log n]$, $k_{j}=I(j,\ell)$ if and only if $k_{j'}=I(j',\ell)~\forall ~j' \in [j]$.
\end{itemize}
\end{obs}
Note that our target is to approximate $\pr\left(k_{\log n}=I(\log n, \ell)=\ell 
\right)$. To do so, $\pr\left(k_{\log n}=\ell\right)$, consider $E_j$, where $j 
\in [\log n] \cup \{0\}$, that denotes the event that $k_j=I(j,\ell)$. 
By Observation~\ref{obs:sampled-all},
\begin{equation}\label{eqn:prob-cond}
 \pr \left( k_{\log n} = \ell \right) = \pr \left( \bigcap\limits_{j=1}^{\log n} 
 E_j\right) = \pr \left(E_0 \right) \prod\limits_{j=1}^{\log n} \pr\left(E_j~\vert~\bigcap\limits_{j'=1}^{j-1} E_{j'} \right)=\prod\limits_{j=1}^{\log n} \pr\left(E_j~\vert~E_{j-1}\right).
\end{equation}
In the above equation, we are using $\pr(E_0)=1$. By the definition of $E_0$, $\pr\left(E_0\right)=\pr \left(k_0=I(0,\ell)\right)$.  Note that our algorithms sets $k_0=1$ deterministically. On the other hand, $I(0,\ell)=1$ as (by Definition~\ref{defi:hset}) $[n]$ is the first and only one hierarchical subset of $[n]$ having label $0$. So, $\pr\left(E_0\right)=1$.

By Equation~\ref{eqn:prob-cond}, we can approximate $\pr(k_{\log n}=\ell)$ by approximating $\pr(E_j~\vert ~E_{j-1})=\pr(k_j=I(j,\ell)~\vert ~k_{j-1}=I(j-1,\ell))$ for each $j\in [\log n]$. For that purpose consider the following observation.
 
\begin{obs}\label{obs:iter-j}
Let $j \in [\log n]$ and $\randsamp(i,\alpha,\delta)$ is executing the $j$-th iteration. Also let $k_{j-1}=k^*$. Then for $k^{\#} \in \{2k^*-1,2k^*\}$, 
$$ (1-4\alpha')
\frac{\ham\left(\mat{A}(i,*)\vert_{S_{j}^{k^{\#}}},\mat{B}(i,*)\vert_{S_{j}^{k^{\#}}}\right)}
{\ham\left(\mat{A}(i,*)\vert_{S_{j-1}^{k^{*}}},\mat{B}(i,*)\vert_{S_{j-1}^{k^{*}}}\right)}\leq \pr \left(k_j=k^{\#}\right) \leq (1+4\alpha')
\frac{\ham\left(\mat{A}(i,*)\vert_{S_{j}^{k^{\#}}},\mat{B}(i,*)\vert_{S_{j}^{k^{\#}}}\right)}
{\ham\left(\mat{A}(i,*)\vert_{S_{j-1}^{k^{*}}},\mat{B}(i,*)\vert_{S_{j-1}^{k^{*}}}\right)}.$$
\end{obs}

\begin{proof}
We prove the stated bound for $\pr(k_j=2k^*-1)$ and the proof for $\pr(k_k=2k^*)$ is similar. In the $j$-th iteration $\randsamp(i,\alpha,\delta)$ calls for $\resdistrows \left(S_{j}
^{2k^*-1},i,\alpha',\delta'\right)$ and $\resdistrows \left(S_{j}^{2k^*},i,\alpha',\delta',\right)
$. Let $d_1^j$ and $d_2^j$ be the respective outputs. Then the algorithm sets $k_j=2k^*-1$ with probability $\frac{d_1^j}{d_1^j+d_2^j}$ and $k_j=2k^*$ with the remaining probability. 
 Note that $d_1^j$ and $d_2^j$ are $(1\pm \alpha')$-approximation of $\ham\left(\mat{A}(i,*)\vert_{S_{j}^{2k^{*}-1}},\mat{B}(i,*)\vert_{S_{j}^{2k^{*}-1}}\right)$ and $\ham\left(\mat{A}(i,*)\vert_{S_{j}^{2k^{*}}},\mat{B}(i,*)\vert_{S_{j}^{2k^{*}}}\right)$,  respectively. By Observation~\ref{obs:hset} (ii), $S_{j-1}^{k^*}=S_{j}^{2k^*-1} \sqcup S_{j-1}^{2k^*}$. So, 
$$ \ham\left(\mat{A}(i,*)\vert_{S_{j-1}^{k^{*}}},\mat{B}(i,*)\vert_{S_{j-1}^{k^{*}}}\right)=\ham\left(\mat{A}(i,*)\vert_{S_{j}^{2k^{*}-1}},\mat{B}(i,*)\vert_{S_{j}^{2k^{*}-1}}\right)+\ham\left(\mat{A}(i,*)\vert_{S_{j}^{2k^{*}}},\mat{B}(i,*)\vert_{S_{j}^{2k^{*}}}\right).$$
 So, $d_1^j + d_2^j$ is a $(1\pm 2 \alpha')$-approximation to $\ham\left(\mat{A}(i,*)\vert_{S_{j-1}^{k^{*}}},\mat{B}(i,*)\vert_{S_{j-1}^{k^{*}}}\right)$. Hence, $\pr(k_j=2k^*-1)=\frac{d_1^j}{d_1^j+d_2^j}$ can be bounded as
 
\begin{eqnarray*}
\small
&& \left(\frac{1-\alpha'}{1+2\alpha'}\right)\frac{\ham\left(\mat{A}(i,*)\vert_{S_{j}^{k^{\#}}},\mat{B}(i,*)\vert_{S_{j}^{k^{\#}}}\right)}
{\ham\left(\mat{A}(i,*)\vert_{S_{j-1}^{k^{*}}},\mat{B}(i,*)\vert_{S_{j-1}^{k^{*}}}\right)} \leq \pr(k_j=2k^*-1) \leq \left(\frac{1+\alpha'}{1-2\alpha'}\right)\frac{\ham\left(\mat{A}(i,*)\vert_{S_{j}^{k^{\#}}},\mat{B}(i,*)\vert_{S_{j}^{k^{\#}}}\right)}{
\ham\left(\mat{A}(i,*)\vert_{S_{j-1}^{k^{*}}},\mat{B}(i,*)\vert_{S_{j-1}^{k^{*}}}\right)}
\end{eqnarray*}
 Note that we are done with the desired bound for $\pr(k_j=2k^*-1)$.
 %From the description of the algorithm, $\pr(k_j=2k^*-1)\frac{d_1^j}{d_1^j+d_2^j}$. Observe that it lies in the interval as given in the statement because of the approximation guarantee for $d_1^j$ and $d_2^j$ as discussed above. Similarly, one can also argue for  $\pr(k_j=2k^*)$.
\end{proof}

For $\pr(E_j~\vert~E_{j-1})$, take $k^*=I(j-1,\ell)$ and $k^{\#}=I(j,\ell)$ in the above Observation. Then we get 
$$
\pr(E_j~\vert ~E_{j-1}) = \pr\left(k_j=I(j,\ell)~\vert ~ k_{j-1}=I(j-1,\ell)\right) \\
\leq (1+4\alpha')\frac{\ham\left(\mat{A}(i,*)\vert_{S_{j}^{I(j,\ell)}}, \mat{B}(i,*)\vert_{S_{j}^{I(j,\ell)}}\right)}{\ham\left(\mat{A}(i,*)\vert_{S_{j-1}^{I(j-1,\ell)}}, \mat{B}(i,*)\vert_{S_{-1j}^{I(j-1,\ell)}}\right)}.
$$
Putting the above for each $j \in [\log n]$ in Equation~\ref{eqn:prob-cond}, we have 

$$\pr(E_j~\vert ~E_{j-1})= \prod\limits_{j=1}^{\log n}(1+4\alpha')\frac{\ham\left(\mat{A}(i,*)\vert_{S_{j}^{I(j,\ell)}}, \mat{B}(i,*)\vert_{S_{j}^{I(j,\ell)}}\right)}{\ham\left(\mat{A}(i,*)\vert_{S_{j-1}^{I(j-1,\ell)}}, \mat{B}(i,*)\vert_{S_{-1j}^{I(j-1,\ell)}}\right)} \leq (1+4\alpha')^{\log n}\frac{1}{\ham\left(\mat{A}(i,*),\mat{B}(i,*)\right)}.$$

Observe that $\ham\left(\mat{A}(i,*),\mat{B}(i,*)\right)=\size{\mbox{{\sc NEQ}(\mat{A},\mat{B},i)}}$. Also recall that $\alpha'=\frac{\alpha}{50 \log n}$. Hence, we have $\pr(E_j~\vert ~E_{j-1})\leq (1+\alpha)\frac{1}{\size{\mbox{{\sc NEQ}(\mat{A},\mat{B},i)}}}$.

The proof of $\pr(E_j~\vert ~E_{j-1})\geq (1-\alpha)\frac{1}{\size{\mbox{{\sc NEQ}(\mat{A},\mat{B},i)}}}$ is similar.

\end{proof}
%%%%%%%%%%%%%%%%%%%%%End of proof of Lemma 2.8 %%%%%%%%%%%%%%%%%%%%%%%%%%%%%%%

%Formal proof of Algorithm 2 ~\ref{algo:random-order} %%%%%%
\section{Formal proof of correctness of Algorithm~\ref{algo:random-order}}
\label{sec:append-formal-proof}
%%%%%%%%%%%%%%%%%%%%% Formal Proof %%%%%%%%%%%%%%%%%%%%%%%    
\noindent
From the above discussion, we need to only prove for the case when $T$ is at least a suitable polynomial in $\log n$ and $\frac{1}{\eps}$. The proof (of correctness) is based on the following lemma that can be proved by mainly using Chernoff bound (See Appendix~\ref{sec:prob}) and some specific details of the algorithm.
% The formal proof of the above lemma is presented in Appendix~\ref{}.
\begin{lem}[Intermediate Lemma to prove correctness of Algorithm~\ref{algo:random-order}]\label{lem:main-inter}
Let $\eps \in \left(0,\frac{1}{2}\right)$, $\beta=\frac{\eps}{50}$ and $\eta=\frac{1}{\poly{n}}$. Consider an oracle $\O_{\beta,\eta}:[n]\rightarrow \R$, as defined in Definition~\ref{defi:orc-apx}, with respect to which Algorithm~\ref{algo:random-order} found $\hat{Y_1},\ldots,\hat{Y_t} \subseteq \Gamma$. Let $Y_1,\ldots,Y_t$ be the buckets into which $[n]$ is partitioned w.r.t. $\O_{\beta,\eta}$. Then, for $k\in [t]$,
\begin{itemize}
\item[(i)] if $\size{Y_k}\geq \frac{\sqrt{\eps T}}{50t}$, then $\pr\left(\size{\frac{n}{\size{\Gamma}}\size{\hat{Y_k}}-\size{Y_k}}\geq \frac{\eps}{50}|Y_k|\right)\leq \frac{1}{\poly{n}}$;
\item[(ii)] if $\size{Y_k}\leq \frac{\sqrt{\eps T}}{50t}
$, then $\pr\left(\frac{n}{\size{\Gamma}}\size{\hat{Y_k}}\geq 
\frac{\sqrt{\eps T}}{40t}\right)\leq \frac{1}{\poly{n}}$;
\item[(iii)] if $\zeta_k \geq \frac{\eps}{50}$, then $\pr\left(\size{\hat{\zeta_k}-\zeta_k}\geq \frac{\eps}{40}\zeta_k\right)\leq \frac{1}{\poly{n}}$;
\item[(iv)] if $\zeta_k \leq \frac{\eps}{50}$, then $\pr\left({\hat{\zeta_k}}\geq \frac{\eps}{30}\right)\leq \frac{1}{\poly{n}}$.
\end{itemize}
\end{lem}
\begin{proof}
\begin{itemize}
\item[(i)] Recall that $\size{\Gamma}=\tOh\left(\frac{n}{\sqrt{T}}\right)$ and $\Gamma \subseteq [n]$ is obtained with replacement. Observe that $\E\left[\size{\hat{Y_k}}\right]=\frac{\size{\Gamma}}{n}\size{Y_k}$. As $\size{Y_k}\geq \frac{\sqrt{\eps T}}{50t}$, $\size{\hat{Y_k}}$ is at least a suitable polynomial in $\log n$ and $\frac{1}{\eps}$. Applying Chernoff bound  (See Lemma~\ref{lem:chernoff} in Appendix~\ref{sec:prob}), we get the desired result.
\item[(ii)] In this case also, $\E\left[\size{\hat{Y_k}}\right]=\frac{\size{\Gamma}}{n}\size{Y_k}$. 
As $\size{Y_k} \leq \frac{\sqrt{\eps T}}{50t}$, $\size{\hat{Y_k}}$ is at most a suitable polynomial in $\log n$ and $\frac{1}{\eps}$.  Applying Chernoff bound  (See Lemma~\ref{lem:chernoff-ul} (i) in Appendix~\ref{sec:prob}), we get the desired result.
\item[(iii)] Recall that 
$\zeta_k=\frac{d_{LS}^k}{\sum\limits_{i \in Y_k}\ham\left(\mat{A}(i,*),\mat{B}(i,*)\right)}$, 
where $d_{LS}^k=\sum\limits_{i\in Y_k} \ham \left(\mat{A}(i,*) \vert_{I_S},\mat{B}(i,*) \vert_{I_S}\right)$. 
By the description of the algorithm, for each $k\in L$, it 
estimates $\zeta_k$ by finding $\hat{\zeta_k}$ as follows. The algorithm determines $Z_k$ of $
\size{\hat{Y_k}}$ many random samples with replacement from $\hat{Y_k}$. Then for each sample $i 
\in Z_k$, it calls for $\randsamp(i,\beta,\eta)$ (where $\beta =\frac{\eps}{50}$) and checks if the output index is in $I_S$, where 
$I_S$ is the set of indices in the small buckets. Let $C_k$ be the number of samples whose 
corresponding call to $\randsamp$ returns an index in $I_S$. The algorithm sets $\hat{\zeta_k}=
\frac{C_k}{\size{\hat{Y_k}}}$.

Observe that while finding $\hat{\zeta_k}$, the algorithm works on a $\hat{Y_k} \subseteq Y_k$ that has been obtained from the random sample $\Gamma \subseteq [n]$. Note that $d_{LS}^k$ can be expressed as 
$$d_{LS}^k=\{(i,j):i\in Y_k~\mbox{and}~j \in I_S~\mbox{such that}~\mat{A}(i,j) \neq \mat{B}(i,j)\}.$$

By the description of $\randsamp(i,\beta,\eta)$ in Lemma~\ref{lem:almost}, observe that, for a fixed $\hat{Y_k}$, 
$$(1-\beta)\sum\limits_{i \in \hat{Y_k}} \ham \left(\mat{A}(i,*)\vert_{I_S},\mat{B}(i,*) \vert_{I_S}\right) \leq \E\left[C_k\right] \leq (1+\beta)\sum\limits_{i \in \hat{Y_k}} \ham \left(\mat{A}(i,*)\vert_{I_S},\mat{B}(i,*) \vert_{I_S}\right).$$

Instead of fixing $\hat{Y_k}$ if we fix $\size{\hat{Y_k}}$, we get 
$$(1-\beta)\frac{\size{\hat{Y_k}}}{\size{Y_k}}\sum\limits_{i \in {Y_k}} \ham \left(\mat{A}(i,*)\vert_{I_S},\mat{B}(i,*) \vert_{I_S}\right) \leq \E\left[C_k\right] \leq (1+\beta)\frac{\size{\hat{Y_k}}}{\size{Y_k}}\sum\limits_{i \in {Y_k}} \ham \left(\mat{A}(i,*)\vert_{I_S},\mat{B}(i,*) \vert_{I_S}\right).$$

Now by the definition of $\zeta_k$ and putting $\beta=\frac{\eps}{50}$,
$$ \left(1-\frac{\eps}{50}\right)\size{\hat{Y_k}}\zeta_k  \leq \E\left[C_k\right] \leq \left(1+\frac{\eps}{50}\right)\size{\hat{Y_k}}\zeta_k.$$

Note that $k$ is in the set $L$ of large buckets. So, $\size{\hat{Y_k}}\geq \frac{\size{\Gamma}}{n}\frac{\sqrt{\eps T}}{40t}$. Also, note that, we are considering the case $\zeta_k \geq \frac{\eps}{50}$. So, $\size{\hat{Y_k}\zeta_k}$ is at least $\frac{\sqrt{\eps T}}{40t} \frac{\eps}{50}$, which is a suitable polynomial in $\log n$ and $\frac{1}{\eps}$. This is because we are assuming $T$ is at least a suitable polynomial in $\log n$ and $\frac{1}{\eps}$. Applying Chernoff bound (See Lemma~\ref{lem:chernoff-ul} in Appendix~\ref{sec:prob}), we can show that $C_k$ is a $\left(1 \pm \frac{\eps}{40}\right)$-approximation to $\size{\hat{Y_k}}\zeta_k$. Hence, $\hat{\zeta_k}$ is a $\left(1 \pm \frac{\eps}{40}\right)$ approximation $\zeta_k$. 

\item[(iv)] The proof of (iv) is similar to that of (iii).
\end{itemize}

\end{proof}
We prove the correctness of our algorithm $\distsymmmat(\mat{A},\mat{B},\vareps,T)$ (Algorithm~\ref{algo:random-order}) via two claims stated below.
\begin{lem}[{\bf Approximating $d_L$}]\label{cl:apx-dl}
$\left(1-\frac{\eps}{50}\right)d_L\leq\hat{d_L} \leq \left(1+\frac{\eps}{50}\right)d_L$ with high probability.
\end{lem}
\begin{lem}[{\bf Approximating $d_S$}]\label{cl:apx-ds}
$\left(1-\frac{\eps}{15}\right)d_{S}-\frac{\eps T}{1600}-\frac{\eps}{25}d_L \leq \hat{d_S} \leq \left(1+\frac{\eps}{15}\right)d_{S}+\frac{\eps}{25}d_{L}.$ holds with high probability.
\end{lem}
Recall that $\dist \left(\mat{A},\mat{B}\right)=d_L+d_S$. Assuming that the above two claims hold, $\hat{d}=\hat{d_L}+\hat{d_S}$ satisfies $\left(1-\frac{\eps}{10}\right)\dist \left(\mat{A},\mat{B}\right)-\frac{\eps}{1600}T \leq \hat{d}\leq \left(1+\frac{\eps}{10}\right)\dist \left(\mat{A},\mat{B}\right)$ with high probability. Note that $\hat{d}$ satisfies the requirement for an estimate of $\dist{ \left(\mat{A},\mat{B}\right)}$ as stated in Lemma~\ref{lem:matrix-dist-symm-T}. Now, it remains to show Lemma~\ref{cl:apx-dl} and~\ref{cl:apx-ds}. We first prove the following claim that follows from our bucketing scheme and will be used in the proofs of Lemma~\ref{cl:apx-dl} and~\ref{cl:apx-ds}. The following claim establishes the connection between the size of a bucket with the sum of the distances between rows (with indices in the same bucket) of matrices $\mat{A}$ and $\mat{B}$.

\begin{cl}\label{cl:orc-up-low-indi}
$\mbox{$\forall k \in [t]$}, \sum\limits_{i\in Y_k} \ham\left(\mtx{A}({i,*}),\mtx{B}({i,*})\right) \leq  \size{Y_k}\left(1+\frac{\eps}{50}\right)^k \leq \left(1+\frac{\eps}{50}\right) \sum\limits_{i\in Y_k} \ham\left(\mtx{A}({i,*}),\mtx{B}({i,*})\right)$ holds with probability at least $\high$.
\end{cl}
\begin{proof}
 $Y_1,\ldots,Y_t \subseteq [n]$ be the buckets into which $[n]$ is partitioned, where $$Y_k=\{i\in [n]: \left(1+\frac{\eps}{50}\right)^{k-1}\leq \O_{\beta,\eta}(i)< \left(1+\frac{\eps}{50}\right)^{k}\}.$$ 
So, 
\begin{eqnarray}
&&  \mbox{For each $k \in [t]$,} \sum\limits_{i\in Y_k}\O_{\beta,\eta}(i) \leq \size{Y_k}\left(1+\frac{\eps}{50}\right)^k \leq \left(1+\frac{\eps}{50}\right)\sum\limits_{i\in Y_k}\O_{\beta,\eta}(i)\label{eqn:Oracle-up-lw}
\end{eqnarray}

 %\mbox{and}&& \sum\limits_{i\in I_L}\O_{\beta,\eta}(i) \leq \sum\limits_{k \in L} \size{Y_k}\left(1+\frac{\eps}{50}\right)^k \leq \left(1+\frac{\eps}{50}\right)\sum\limits_{i\in I_L}\O_{\beta,\eta} {eqn:Oracle-up-lw}

Here $\beta=\frac{\eps}{50}$ and $\eta=\frac{1}{\poly{n}}$.

Oracle $\O_{\beta,\eta}:[n]\rightarrow \R$ is a function, as defined in Definition~\ref{defi:orc-apx}, such that $\O_{\beta,\eta}(i)$ equals to $\hat{a_i}$, the value returned by $\distrows(i,\beta,\eta)$, if the algorithm $\distrows(i,\beta,\eta)$ is called (once). Otherwise, $\O_{\beta,\eta}(i)$ is set to some $(1\pm \beta)$-approximation to $\ham\left(\mat{A}(i,*),\mat{B}(i,*)\right)$. So, Equation~\ref{eqn:Oracle-up-lw} implies that the following holds with probability at least $1-\frac{1}{\poly{n}}$.
$$  \mbox{$\forall k \in [t]$}, \sum\limits_{i\in Y_k} \ham\left(\mtx{A}({i,*}),\mtx{B}({i,*})\right) \leq  \size{Y_k}\left(1+\frac{\eps}{50}\right)^k \leq \left(1+\frac{\eps}{50}\right) \sum\limits_{i\in Y_k} \ham\left(\mtx{A}({i,*}),\mtx{B}({i,*})\right).
$$
\end{proof}
Now we will show Lemma~\ref{cl:apx-dl}.
\begin{proof}[Proof of Lemma~\ref{cl:apx-dl}]
Note that $d_L=\sum\limits_{i\in I_L} \ham\left(\mtx{A}({i,*}),\mtx{B}({i,*})\right)$ and $\hat{d_L}=\frac{n}{\size{\Gamma}}\sum\limits_{k\in L}\size{\hat{Y_k}}\left(1+\frac{\eps}{50}\right)^k$.

By the definition of $d_L$ as well as Claim~\ref{cl:orc-up-low-indi}, we get
\begin{equation}\label{eqn:d_l-buck-orc}
\pr\left(d_L \leq \sum\limits_{k \in L} \size{Y_k}\left(1+\frac{\eps}{50}\right)^k \leq \left(1+\frac{\eps}{50}\right)d_L \right) \geq 1-\frac{1}{\poly{n}}.
\end{equation}

Recall that, for $k\in [t]$ in the set $L$ of large buckets, $\hat{Y_k} \geq \tau$. Here $\tau= \frac{\size{\Gamma}}{n}\frac{\sqrt{\vareps T}}{40 t}$. By Lemma~\ref{lem:main-inter} (ii) and (i), for each $k \in L$, $\frac{n}{\size{\Gamma}}\size{\hat{Y_k}}$ is an $\left(1\pm \frac{\eps}{50}\right)$-approximation to $\size{Y_k}$ with probability at least $1-\frac{1}{\poly{n}}$. So, the following holds with probability at least $1-\frac{1}{\poly{n}}$.
\begin{eqnarray}
&& \left(1-\frac{\eps}{50}\right)d_L \leq \frac{n}{\size{\Gamma}}\sum\limits_{k\in L}\size{\hat{Y_k}}\left(1+\frac{\eps}{50}\right)^k \leq \left(1+\frac{\eps}{50}\right)^2d_L \label{eqn:apx-dl}
\end{eqnarray}

By the definition of $\hat{d_L}$ along with taking $\eps \in \left(0,\frac{1}{2}\right)$, we conclude that 
\begin{eqnarray*}
&&\pr\left(\left(1-\frac{\eps}{20}\right)d_L \leq \hat{d_L} \leq \left(1+\frac{\eps}{20}\right)d_L \right) \geq 1-\frac{1}{\poly{n}}.
\end{eqnarray*}
%holds with probability $1-\frac{1}{\poly{n}}$.
\end{proof}

\begin{proof}[Proof of Lemma~\ref{cl:apx-ds}]
Recall that $d_S=\sum\limits_{i\in I_S} \ham\left(\mtx{A}({i,*}),\mtx{B}({i,*})\right)$ and $\hat{d_S}=\frac{n}{\size{\Gamma}}\sum\limits_{k\in L}\hat{\zeta_k}\left(1+\frac{\eps}{50}\right)^k$. Moreover, $d_S=d_{SL}+d_{SS}$, where $d_{SL}=\sum\limits_{i\in I_S} \ham \left(\mat{A}(i,*)\vert_{I_L},
\mat{B}(i,*)\vert_{I_L}\right)$ and $d_{SS}=\sum\limits_{i\in I_S} \ham \left(\mat{A}(i,*)\vert_{I_S},
\mat{B}(i,*)\vert_{I_S}\right)$.  Also, as $\mat{A}$ and $\mat{B}$ are symmetric 
matrices, $d_{SL}=d_{LS}=\sum\limits_{i\in I_L} \ham\left(\mat{A}(i,*)
\vert_{I_S},\mat{B}(i,*)\vert_{I_S}\right)$. So, $d_S=d_{SS}+d_{LS}$.

First we show that 
\begin{cl}\label{cl:bound:dss}
$\pr\left(d_{SS} \leq \frac{\eps T}{1600}\right) \geq \high.$
\end{cl}
Then we show that
\begin{cl}\label{cl:bound-dsl}
$\pr\left(\left(1-\frac{\eps}{15}\right)d_{LS}-\frac{\eps}{25}d_L  \leq \hat{d_S} \leq \left(1+\frac{\eps}{15}\right)d_{LS} + \frac{\eps}{25}d_L \right) \geq \high.$
\end{cl}
As $d_S=d_{SS}+d_{LS}$, the above two claims imply the Lemma, that is, the following holds with probability at least $\high$.
$$\left(1-\frac{\eps}{15}\right)d_{S}-\frac{\eps T}{1600}-\frac{\eps}{25}d_L \leq \hat{d_S} \leq \left(1+\frac{\eps}{15}\right)d_{S}+\frac{\eps}{25}d_{L}.$$
So, it remains to show Claims~\ref{cl:bound:dss} and~\ref{cl:bound-dsl}.
\begin{proof}[Proof of Claim~\ref{cl:bound:dss}]
Note that $d_{SS}=\sum\limits_{i\in I_S} \ham \left(\mat{A}(i,*)\vert_{I_S},
\mat{B}(i,*)\vert_{I_S}\right)=\size{\{(i,j) \in I_S \times I_S~\mbox{:}~ \mat{A}(i,j)\neq \mat{B}(i,j)\}} $, where $I_S$ denotes the set of indices in the $Y_k$'s with $k \in S$ and $S$ is the set of small buckets. So, $\size{I_S}=\sum\limits_{k \in S}\size{Y_k}$. By the definition of $S$, for every $k \in S$, $\hat{Y_k} < \tau = 
\frac{\size{\Gamma}}{n}\frac{\sqrt{\eps T}}{50 t}$. By Lemma~\ref{lem:main-inter} 
(i), we have $\size{Y_k} \leq \left(1+\frac{\eps}{50} \right) \frac{n}{\
\size{\Gamma}}\size{\hat{Y_k}} \leq \frac{\sqrt{\eps T}}{40 t}$ with 
probability at least $\high$. This implies that $\size{I_S}=\sum\limits_{k \in S}
\size{Y_k} \leq \frac{\sqrt{\eps T}}{40}$  with 
probability at least $\high$. Hence, by the definition of $d_{SS}$, we have the following with probability at least $\high$.
$$d_{SS}=\sum\limits_{i\in I_S} \ham\left(\mtx{A}({i,*}),\mtx{B}({i,*})\right) \leq \size{I_S}^2 \leq \frac{\eps T}{1600}.$$
\end{proof}
\begin{proof}[{Proof of claim~\ref{cl:bound-dsl}}]
Note that $d_{LS}=\sum\limits_{i\in I_L} \ham \left(\mat{A}(i,*)\vert_{I_S},
\mat{B}(i,*)\vert_{I_S}\right)$. Recall that $d_{LS}=\sum\limits_{k \in L}=d_{LS}^k$, where $d_{LS}^k=\sum\limits_{i \in Y_k} \ham\left(\mat{A}(i,*)
\vert_{I_S},\mat{B}(i,*)\vert_{I_S}\right).$ Also, recall that 
$\zeta_k=\frac{d_{LS}^k}{\sum\limits_{k \in L}\ham\left(\mat{A}(i,*)
,\mat{B}(i,*)\right)}$. So, $d_{SL}^k=\zeta_k \sum \limits_{i \in Y_k}\ham\left(\mat{A}(i,*)
,\mat{B}(i,*)\right)$. Hence,
\begin{equation}\label{eqn:dsl-exact}
d_{LS}=\sum\limits_{k \in L} \zeta_k \sum\limits_{i \in Y_k} \ham\left(\mat{A}(i,*)
,\mat{B}(i,*)\right).
\end{equation}
By Claim~\ref{cl:orc-up-low-indi}, the following holds with probability at least $\high$.
$$\forall k\in [t], \zeta_k\sum\limits_{i\in Y_k} \ham\left(\mtx{A}({i,*}),\mtx{B}({i,*})\right) \leq  \zeta_k \size{Y_k}\left(1+\frac{\eps}{50}\right)^k \leq \zeta_k \left(1+\frac{\eps}{50}\right) \sum\limits_{i\in Y_k} \ham\left(\mtx{A}({i,*}),\mtx{B}({i,*})\right) ). $$
Taking sum over all $k \in L$ and then applying Equation~\ref{eqn:dsl-exact}, the following holds with probability at least $\high$.
$$\pr\left( d_{LS} \leq \sum\limits_{k \in L} \zeta_k \size{Y_k}\left(1+\frac{\eps}{50}\right)^k  \leq \left(1+\frac{\eps}{50}\right)d_{LS} \right) \geq \high .$$
Recall that, for $k\in [t]$ in the set $L$ of large buckets, $\hat{Y_k} \geq \tau$. Here $\tau= \frac{\size{\Gamma}}{n}\frac{\sqrt{\vareps T}}{40 t}$. By Lemma~\ref{lem:main-inter} (ii) and (i), for each $k \in L$, $\frac{n}{\size{\Gamma}}\size{\hat{Y_k}}$ is a $\left(1\pm \frac{\eps}{50}\right)$-approximation to $\size{Y_k}$ with probability at least $1-\frac{1}{\poly{n}}$. So,
\begin{equation}\label{eqn:dsl-apx-1}
\pr\left( \left(1-\frac{\eps}{50}\right) d_{LS} \leq \frac{n}{\size{\Gamma}}\sum\limits_{k \in L} \zeta_k \size{\hat{Y_k}}\left(1+\frac{\eps}{50}\right)^k  \leq \left(1+\frac{\eps}{50}\right)^2d_{LS} \right) \geq \high .
\end{equation} 
Having the above equation, consider $\hat{d_S}=\frac{n}{\size{\Gamma}}\sum\limits_{k \in L} \hat{\zeta_k} \size{\hat{Y_k}}\left(1+\frac{\eps}{50}\right)^k$ whose upper and lower bound is to be proved as stated in Claim~\ref{cl:bound-dsl}. Breaking the sum into two parts depending the values of $\hat{\zeta_k}$'s, we have 
\begin{equation}\label{eqn:dsl-break}
\hat{d_S}=\frac{n}{\size{\Gamma}}\sum\limits_{k \in L : \zeta_k \geq \frac{\eps}{50}} \hat{\zeta_k} \size{\hat{Y_k}}\left(1+\frac{\eps}{50}\right)^k +\frac{n}{\size{\Gamma}}\sum\limits_{k \in L :\zeta_k <\frac{\eps}{50}} \hat{\zeta_k} \size{\hat{Y_k}}\left(1+\frac{\eps}{50}\right)^k.
\end{equation}

We prove the desired upper and lower bound on $\hat{d_S}$ separately by using the following observation about upper and lower bounds of the two terms in Equation~\ref{eqn:dsl-break}.
\begin{obs}\label{obs:inter-dsl}
\begin{itemize}
\item[(i)] $\frac{n}{\size{\Gamma}}\sum\limits_{k \in L : \zeta_k \geq \frac{\eps}{50}} \hat{\zeta_k} \size{\hat{Y_k}}\left(1+\frac{\eps}{50}\right)^k \leq \left(1 + \frac{\eps}{15}\right)d_{LS}$ with probability $\high$.
\item[(ii)] $\frac{n}{\size{\Gamma}}\sum\limits_{k \in L :\zeta_k <\frac{\eps}{50}} \hat{\zeta_k} \size{\hat{Y_k}}\left(1+\frac{\eps}{25}\right)^k \leq \frac{\eps}{35}d_{L}$ with probability at least $\high$ .
\item[(iii)] $\frac{n}{\size{\Gamma}}\sum\limits_{k \in L : \zeta_k \geq \frac{\eps}{50}} \hat{\zeta_k} \size{\hat{Y_k}}\left(1+\frac{\eps}{50}\right)^k \geq \left(1 - \frac{\eps}{15}\right)d_{LS}-\frac{\eps}{25}d_L$ with probability $\high$.
\end{itemize}
\end{obs}
\begin{proof} 
\begin{itemize}
\item[(i)] By Lemma~\ref{lem:main-inter} (iii), for each $k \in [t]$ with $\zeta_k \geq \frac{\eps}{50}$, $\hat{\zeta_k}$ is a $\left(1\pm \frac{\eps}{40}\right)$-approximation to $\zeta_k$ with probability at least $\high$. So, with probability at least $\high$, we can derive the following.
\begin{eqnarray*}
\frac{n}{\size{\Gamma}}\sum\limits_{k \in L : \zeta_k \geq \frac{\eps}{50}} \hat{\zeta_k} \size{\hat{Y_k}}\left(1+\frac{\eps}{40}\right)^k &\leq& \left(1 + \frac{\eps}{40}\right)\frac{n}{\size{\Gamma}}\sum\limits_{k \in L : \zeta_k \geq \frac{\eps}{50}} {\zeta_k} \size{\hat{Y_k}}\left(1+\frac{\eps}{50}\right)^k.\\
&\leq& \left(1+\frac{\eps}{40}\right) \left( 1+\frac{\eps}{50}\right)^2 d_{LS} ~\left(\because \mbox{By Equation~\ref{eqn:dsl-apx-1}}\right) \\
&\leq& \left(1+\frac{\eps}{15}\right)d_{LS}.
\end{eqnarray*}
%As $\eps \in \left(0,\frac{1}{2}\right)$, combining the above expression with Equation~\ref{eqn:dsl-apx-1}, we are done with the proof of (i).

\item[(ii)] By Lemma~\ref{lem:main-inter} (iv), for each $k \in [t]$ with $\zeta_k < \frac{\eps}{50}$, $\hat{\zeta_k}$ is at most $\frac{\eps}{30}$ with probability at least $\high$. Hence, the following derivations hold with probability $\high$.
\begin{eqnarray*}
 \frac{n}{{\size{\Gamma}}}\sum\limits_{k \in L : \zeta_k \geq \frac{\eps}{50}} {\zeta_k} \size{\hat{Y_k}}\left(1+\frac{\eps}{50}\right)^k &\leq& \frac{\eps}{30}\cdot \frac{n}{\size{\Gamma}}\sum\limits_{k \in L :\zeta_k <\frac{\eps}{50}} \size{\hat{Y_k}} \left(1+\frac{\eps}{50}\right)^k \\
&\leq& \frac{\eps}{30}\cdot \frac{n}{\size{\Gamma}}\sum\limits_{k \in L } \size{\hat{Y_k}} \left(1+\frac{\eps}{50}\right)^k \\
&\leq& \frac{\eps}{30} \left(1+\frac{\eps}{50}\right)^2 d_{L}~~\left(\mbox{ By Equation~\ref{eqn:apx-dl}}\right)\\
&\leq& \frac{\eps}{25}d_L
\end{eqnarray*}
%Now by Equation~\ref{eqn:apx-dl}, $\frac{n}{\size{\Gamma}}\sum\limits_{k \in L } \size{\hat{Y_k}} \left(1+\frac{\eps}{50}\right)^k \leq \left(1+\frac{\eps}{50}\right)^2d_L$ with probability $\high$. So, we are done with the proof of (ii) as $\eps \in \left(0,\frac{1}{2}\right)$.

\item[(iii)] Note that 
$$\frac{n}{\size{\Gamma}}\sum\limits_{k \in L : \zeta_k \geq \frac{\eps}{50}} \hat{\zeta_k} \size{\hat{Y_k}}\left(1+\frac{\eps}{50}\right)^k \geq \frac{n}{\size{\Gamma}}\sum\limits_{k \in L} \hat{\zeta_k} \size{\hat{Y_k}}\left(1+\frac{\eps}{50}\right)^k -\frac{n}{\size{\Gamma}}\sum\limits_{k \in L : \zeta_k \leq \frac{\eps}{50}} \hat{\zeta_k} \size{\hat{Y_k}}\left(1+\frac{\eps}{50}\right)^k .$$
By Lemma~\ref{lem:main-inter} (iii), for each $k \in [t]$ with $\zeta_k \geq \frac{\eps}{50}$, $\hat{\zeta_k}$ is a $\left(1\pm \frac{\eps}{40}\right)$-approximation to $\zeta_k$ with probability at least $\high$. Also, by Observation~\ref{obs:inter-dsl} (iii), $\frac{n}{\size{\Gamma}}\sum\limits_{k \in L : \zeta_k \leq \frac{\eps}{50}} \hat{\zeta_k} \size{\hat{Y_k}}\left(1+\frac{\eps}{50}\right)^k \leq \frac{\eps}{25}d_L$ with probability at least $\high$. So, we can derive the following with probability at least $\high$.
\begin{eqnarray*}
\frac{n}{\size{\Gamma}}\sum\limits_{k \in L : \zeta_k \geq \frac{\eps}{50}} \hat{\zeta_k} \size{\hat{Y_k}}\left(1+\frac{\eps}{50}\right)^k &\geq& \left(1-\frac{\eps}{40}\right)\frac{n}{\size{\Gamma}}\sum\limits_{k \in L} {\zeta_k} \size{\hat{Y_k}}\left(1+\frac{\eps}{50}\right)^k -\frac{\eps}{25}d_L \\
&\geq& \left(1-\frac{\eps}{40}\right)^2d_{LS}-\frac{\eps}{25}d_L ~\left(\mbox{By Equation~\ref{eqn:dsl-apx-1}}\right)\\
&\geq& \left(1-\frac{\eps}{15}\right)d_{LS}-\frac{\eps}{25}d_L.
\end{eqnarray*}
\end{itemize} 
 
\end{proof}
\remove{By Lemma~\ref{lem:main-inter} (iii), for each $k \in [t]$ with $\zeta_k \geq \frac{\eps}{2}$, $\hat{\zeta_k}$ is a $\left(1\pm \frac{\eps}{50}\right)$-approximation to $\zeta_k$ with probability at least $\high$. So, we can upper and lower bound the first term in the RHS of Equation~\ref{eqn:dsl-break} as follows. \begin{equation}\label{eqn:dsl-first-term}
\Pr\left(\left(1-\frac{\eps}{50}\right) \frac{n}{\size{\Gamma}} \sum\limits_{k \in L : \zeta_k \geq \frac{\eps}{50}} \zeta_k \size{\hat{Y_k}}\left(1+\frac{\eps}{50}\right)^k\leq \frac{n}{\size{\Gamma}}\sum\limits_{k \in L : \zeta_k \geq \frac{\eps}{50}} \hat{\zeta_k} \size{\hat{Y_k}}\left(1+\frac{\eps}{50}\right)^k \leq \left(1+\frac{\eps}{50}\right) \frac{n}{\size{\Gamma}} \sum\limits_{k \in L : \zeta_k \geq \frac{\eps}{50}} \zeta_k \size{\hat{Y_k}}\left(1+\frac{\eps}{50}\right)^k\right) \geq \high.
\end{equation}}
Considering the expression for $\hat{d_S}$ in Equation~\ref{eqn:dsl-break} along with Observation~\ref{obs:inter-dsl} (i) and (ii), we can derive the desired upper bound on $\hat{d_{S}}$ (as follows) that holds with probability at least $\high$.
$$\hat{d_S} \leq \left(1+\frac{\eps}{15}\right)d_{LS} +\frac{\eps}{25}d_L.$$
 For the lower bound part of $\hat{d_S}$, again consider the expression for $\hat{d_S}$ in Equation~\ref{eqn:dsl-break} along with Observation~\ref{obs:inter-dsl} (iii). We have the following with probability at least $\high$.
 $$\hat{d_S} \geq  \left(1-\frac{\eps}{15}\right)d_{LS}-\frac{\eps}{25}d_L.$$
 
\remove{ Recall that $d_S=d_{SL}+d_{SS}$, $d_{SL}=d_{SS}$. Also, we have shown that $d_{SS}\leq \frac{\eps T}{400}$ holds with high probability, we have the following lower bound on $d_S$ with probability $\high$.
 
 $$\hat{d_S} \geq  \left(1-\frac{\eps}{50}\right)\left(d_{S}-\frac{\eps T}{400}\right).$$}
 \end{proof}
\end{proof}
%%%%%%%%%%%%%%%%%%%%End of Formal proof %%%%%%%%%%%%%%%%%%%%%%%%%%
\section{Proof of Lemma~\ref{lem:matrix-dist-symm}}\label{sec:pf-symm}
\noindent
Recall $\distsymmmatguess(\mat{A},\mat{B},\vareps,T)$ and Lemma~\ref{lem:matrix-dist-symm-T} --- $\distsymmmatguess(\mat{A},\mat{B},\vareps,T)$ outputs a $(1\pm \eps)$-approximation to $\dist(\mtx{A},\mtx{B})$ with high probability when the guess $T \leq \dist(\mtx{A},\mtx{B})$ and $T$ (and hence $\dist(\mat{A},\mat{B})$) is at least a suitable polynomial in $\log n$ and $1/\eps$, say $\psi$.

In this Section, we prove Lemma~\ref{lem:matrix-dist-symm} by providing algorithm $\distsymmmat(\mat{A},\mat{B},\eps)$ (Algorithm~\ref{algo:dist-wlow}) to determine a $(1\pm \eps)$ approximation to $\dist(\mat{A},\mat{B})$, with high probability. Note that $\distsymmmat(\mat{A},\mat{B},\eps)$ does not need any known non-trivial lower bound $T$ and it works for all $\dist(\mat{A},\mat{B})$ with $0 \leq \dist(\dist{A},\dist{B}) \leq n^2$. $\distsymmmat(\mat{A},\mat{B},\eps)$ calls $\distsymmmatguess(\mat{A},\mat{B},\eps,T)$  recursively at most $\Oh(\log n)$ times for different values of $T$ until we have {$T \leq \frac{\widehat{m_T}}{1+{\eps}/{10}}$}, where $\hat{m_T}$ denotes the output of $\distsymmmatguess$ $(\mat{A},\mat{B},\eps,T)$. 

Moreover, $\distsymmmat(\mat{A},\mat{B},\eps)$ does not call $\distsymmmatguess(\mat{A},\mat{B},\eps,T)$ if $T$ becomes less than $\psi$ at some point of time. In that situation,  $\distsymmmat(\mat{A},\mat{B},\eps)$ finds a $(1 \pm \eps)$-approximation to $\dist(\mat{A},\mat{B})$ by using the algorithm  described in the following remark.
\begin{rem}\label{rem:triv-algo}
Recall algorithm $\distrows (i,\alpha,\delta)$ as stated in Lemma~\ref{lem:distrows}. It takes $i\in [n]$ and $\alpha ,\delta\in (0,1)$ as inputs, and reports a $(1\pm \alpha)$-approximation 
 to $\ham\left(\mtx{A}({i,*}),\mtx{B}({i,*})\right)$  with 
 probability at least $1-\delta$, and makes $\Oh\left(\frac{\log n}{\alpha^2}  \log 
 \frac{1}{\delta}\right)$ \iporacle queries to both $\mtx{A}$ and $\mtx{B}$.
{Consider the following trivial algorithm, that finds $a_i=\distrows (i,\alpha,\delta)$ by setting $\alpha=\eps$ and $\delta=\frac{1}{n^{10}}$,  for each $i \in [n]$, and gives outputs $\hat{d}=\sum_{i=1}^na_i$ as a $(1 \pm \eps)$-approximation of $\dist({\mat{A},\mat{B}})$. The correctness of the algorithm follows from Observations~\ref{obs:dist} and Lemma~\ref{lem:distrows}. The query complexity of the algorithm is $\tOh(n)$ by Lemma~\ref{lem:distrows}. }
\end{rem}

\remove{Also, note that Lemma~\ref{lem:matrix-dist-symm-T} gives the stated approximation with high probability when $T$ is more than a suitable polynomial in $\log n$ and ${1}/{\eps}$. \comments{To take care of smaller $T$, we would like to note that there is trivial algorithm that makes $\tOh(n)$ \iporacle{} queries and returns a $(1 \pm \eps)$-approximation of $\dist(\mat{A},\mat{B})$ with high probability. The algorithm will be discussed in in Remark~\ref{rem:triv-algo} in Section~\ref{sec:prelim-tech}, and it is good enough for our purpose when $T$ is at least a polynomial in $\log n$ and $1/\eps$. }
}

 \begin{algorithm}
%\scriptsize
\caption{$\distsymmmat(\mat{A},\mat{B},\eps)$}
\label{algo:dist-wlow}
\KwIn{$\epsilon \in (0,1)$.} 
\KwOut{$\widehat{m}$, which is an $(1 \pm \eps)$-approximation of
$\dist(\mat{A},\mat{B})$.}
%\LinesNumbered
\Begin
	{
Initialize $T=n^2/2$.

\While {($T \geq \psi$)}
{
Call $\distsymmmat(\eps,T)$ (Algorithm~\ref{algo:random-order}) and let $\widehat{m_T}$ be the output.

\If{$\left(T \leq \frac{\widehat{m_T}}{1+{\eps}/{10}}\right)$}
{Report $\widehat{m_{T}}$ as the output and {\sc Quit}.}
\Else
{ 
Set $T$ as $T/2$ and {\sc Continue}.
}
}

Call the algorithm discussed in Remark~\ref{rem:triv-algo} as reports its output as the output of  $\distsymmmat(\mat{A},\mat{B},\eps)$.
}
\end{algorithm}

We will argue the correctness of $\distsymmmat(\mat{A},\mat{B},\eps)$ (Algorithm~\ref{algo:dist-wlow}) via the following two observations. 
\begin{obs}\label{obs:obs1}
 $\distsymmmat(\mat{A},\mat{B},\eps)$ does not {\sc Quit} as long as $T>\dist{(\mat{A},\mat{B})}$ and $T \geq \psi$, with high probability.
\end{obs}
\begin{proof}
Consider a fixed $T$ with $T>\dist(\mat{A},\mat{B})$ and $T \geq \psi$. By Lemma~\ref{lem:matrix-dist-symm-T},  $\widehat{m_T} \leq \left(1+\eps/10\right)\dist{(\mat{A},\mat{B})}$ with high probability. As $T > \dist{(\mat{A},\mat{B})}$, $\widehat{m_T} < \left(1+\eps/10\right)T$ with high probability. This implies $T >\frac{\widehat{m_T}}{1+\eps/10} $, that is, $\distsymmmat(\mat{A},\mat{B},\eps)$ does not {\sc Quit} for this fixed $T$ with high probability. As there can be at most $\Oh(\log n)$ many $T$'s with $T > \dist{(\mat{A},\mat{B})}$ such that $\distsymmmat(\mat{A},\mat{B},\eps)$ calls $\distsymmmatguess(\mat{A},\mat{B},\eps,T)$, we are done with the proof. 
\end{proof}
\begin{obs}\label{obs:obs2}
Let $\psi \leq T \leq \dist{(\mat{A},\mat{B})}/30$. $\distsymmmat(\mat{A},\mat{B},\eps)$ quits and reports $\widehat{m_{T}}$ as the output with high probability.
\end{obs}
\begin{proof}
By Claim~\ref{lem:matrix-dist-symm-T}, with high probability, $\left(1-\frac{\eps}{10}\right) \dist{(\mat{A},\mat{B})}-\frac{\eps T}{1600} \leq \widehat{m_T}\leq  \left(1+\frac{\eps}{10}\right)\dist{(\mat{A},\mat{B})}$. As $\psi \leq T \leq \frac{\dist{\left(\mat{A},\mat{B}\right)}}{30}$, with high probability, we have $\left(1-{\eps}\right) \dist{(\mat{A},\mat{B})} \leq \widehat{m_T}\leq  \left(1+\eps \right) \dist{(\mat{A},\mat{B})}$. So, $T \leq \frac{\widehat{m_T}}{30(1-\eps)}$ with high probability. As $\eps \in \left(0,\frac{1}{2}\right)$, we have  $T \leq \frac{\widehat{m_T}}{1+\eps/10}$ with high probability. Hence, $\distsymmmat(\mat{A},\mat{B},\eps)$ quits and reports a $(1 \pm \eps)$-approximation to $\dist{(\mat{A},\mat{B})}$ (as the output) with high probability.
\end{proof}
Now, we analyze the correctness of $\distsymmmat(\mat{A},\mat{B},\eps)$ by dividing into two cases:
\begin{description}
\item[$\dist(\mat{A},\mat{B})\geq 60 \psi$:] From Observations~\ref{obs:obs1} and~\ref{obs:obs2}, $\distsymmmat(\mat{A},
\mat{B},\eps)$ does not quit when the function $\distsymmmatguess(\mat{A},
\mat{B},\eps,T)$ is called for any $T > \dist(\mat{A},\mat{B})$ and quits when $
\distsymmmatguess(\mat{A},\mat{B},\eps,T)$ is called for some $ 
\frac{\dist{(\mat{A},\mat{B})}}{60}\leq T \leq \dist{(\mat{A},\mat{B})}$, with 
high probability. So, as $\dist(\mat{A},\mat{B})\geq 60 \psi$, $T$ does not go below $\psi$ with high probability. By the description of $\distsymmmat(\mat{A},
\mat{B},\eps)$, it reports the output by the last call to  $\distsymmmatguess$ $(\mat{A},
\mat{B},\eps,T)$ with $\psi \leq \frac{\dist{(\mat{A},\mat{B})}}{60} \leq T \leq \dist{(\mat{A},\mat{B})}$. By Lemma~\ref{lem:matrix-dist-symm-T}, the output returned by $\distsymmmat(\mat{A},
\mat{B},\eps)$ is a $(1 \pm \eps)$-approximation to $\dist\left(\mat{A},\mat{B}\right)$. Observe that the 
query complexity of the last call to $\distsymmmatguess(\mat{A},\mat{B},\eps,T)$ is 
$\tOh\left({n}/{\sqrt{\dist{(\mat{A},\mat{B})}}}\right)$. For the total query complexity of $\distsymmmat(\mat{A},
\mat{B},\eps)$, note that $\distsymmmat(\mat{A},\mat{B},\eps)$ reduces $T$ by a 
factor of $2$ each time it does not quit. So, $\distsymmmatguess(\mat{A},\mat{B},
\eps,T)$ is called $\Oh(\log n)$ times. Observe that the 
query complexity of $\distsymmmat(\mat{A},\mat{B},\eps)$ is dominated by the 
query complexity of the last call to $\distsymmmatguess(\mat{A},\mat{B},\eps,T)
$. Hence, the total number of queries made by $
\distsymmmat(\mat{A},\mat{B},\eps)$ is 
$\tOh\left({n}/{\sqrt{\dist{(\mat{A},\mat{B})}}}\right)$.
\item[$\dist(\mat{A},\mat{B})< 60 \psi$:] From Observation~\ref{obs:obs1}, $\distsymmmat(\mat{A},
\mat{B},\eps)$ does not quit when the function $\distsymmmatguess(\mat{A},
\mat{B},\eps,T)$ is called for any $T > \dist(\mat{A},\mat{B})$. So, as $\dist(\mat{A},\mat{B})< 60 \psi$, $T$ goes below $60\psi$ with high probability, and $T$ may go below $\psi$. If $T$ does not go below $\psi$, the analysis is same as that of the case when $\dist(\mat{A},\mat{B})\geq 60 \psi$.  Now we analyze the correctness of $\distsymmmat(\mat{A},\mat{B},\eps)$ and query complexity assuming that $T$ becomes less than $\psi$ at some point of time. In this case,  $\distsymmmat(\mat{A},\mat{B},\eps)$ calls the algorithm discussed in Remark~\ref{rem:triv-algo}, and reports its output as the output of  $\distsymmmat(\mat{A},\mat{B},\eps)$. The number of \iporacle queries performed by $\distsymmmat(\mat{A},\mat{B},\eps)$ here is $\tOh(n)$, which is $\tOh\left(n/\dist(\mat{A},\mat{B})\right)$ as   $\dist(\mat{A},\mat{B})< 60 \psi$. The required approximation guarantee for $\dist(\mat{A},\mat{B})$ follows directly from Remark~\ref{rem:triv-algo}. For the total query complexity of $\distsymmmat(\mat{A},\mat{B},\eps)$, observe that it is dominated by the number of \iporacle queries performed by the algorithm corresponding to Remark~\ref{rem:triv-algo}.
\end{description}

So, we are done with the proof of Lemma~\ref{lem:matrix-dist-symm}.

\remove{\section{JL Lemma}\label{sec:JL}

\begin{lem}\label{lem:JL}[Johnson–Lindenstrauss Lemma]
Let us consider any pair of points ${\bf u}, {\bf v} \in \R^N$. For a given $\eps \in (0,1)$ and $\delta \in (0,1)$, there is a map $f:\R^N \rightarrow \R^d$ such that $d=\Theta\left(\frac{1}{\eps^2}\log \frac{1}{\delta}\right)$ satisfying the following.
\begin{equation}\label{eq:JL}
 (1-\eps)||{\bf u}-{\bf v}||_2^2 \leq  ||f({\bf u})-f({\bf v})||_2^2 \leq (1+\eps)||u-v||_2^2.
\end{equation}
\end{lem}}

\section{Probability Results}
\label{sec:prob}

\begin{lem} [See~\cite{Dubhashi09}]
\label{lem:chernoff}
Let $X = \sum_{i \in [n]} X_i$ where $X_i$, $i \in [n]$, are independent random variables,  $X_i \in [0,1]$ and $\mathbb{E}[X]$ is the expected value of $X$. Then
 for $\epsilon \in (0,1)$,
	$\Pr \left[ \left|X - \mathbb{E}[X] \right| > \epsilon \mathbb{E} \left[ X \right] \right] \leq \exp{ \left(- \frac{\epsilon^2}{3} \mathbb{E}[X]\right)}$.
 
\end{lem}

\begin{lem} [See~\cite{Dubhashi09}]
\label{lem:chernoff-ul}
Let $X = \sum_{i \in [n]} X_i$ where $X_i$, $i \in [n]$, are independent random variables,  $X_i \in [0,1]$ and $\mathbb{E}[X]$ is the expected value of $X$. Suppose $\mu_L \leq \mathbb{E}[X] \leq \mu_H$, then for $0 < \epsilon < 1$,
\begin{itemize}
\item[(i)] $\Pr[ X > (1+\epsilon)\mu_H] \leq \exp{\left( - \frac{\epsilon^2}{3} \mu_H\right)}$.
\item[(ii)] $\Pr[ X < (1-\epsilon)\mu_L] \leq \exp{\left( - \frac{\epsilon^2}{2} \mu_L\right)}$.
\end{itemize}
\end{lem}

\remove{\item[(ii)] \label{ch02} if $t > 2 \mathbb{E}[X]$, then 
\begin{eqnarray}
	\Pr[ X > t ] \leq 2^{-t}
\end{eqnarray}}

\end{document}